\documentclass[sigconf]{acmart}

\usepackage{bm}
\usepackage{multirow}
\usepackage{enumitem}
\usepackage{algorithm}
\usepackage{algorithmicx}
\usepackage{algpseudocode}
\AtBeginDocument{%
  }

\setcopyright{none}

\acmConference[arXiv '26]{arXiv}{April, 2026}{}

\begin{document}
\title{SurgGraph: Quantitative Laparoscopic Video Understanding via Geometry-Grounded Scene Graphs}
\author{Jingying Wang}
\affiliation{%
  \institution{University of Michigan}
  \city{Ann Arbor}
  \state{Michigan}
  \country{United States}
}
\email{wangchy@umich.edu}

\author{Rosiana Natalie}
\affiliation{%
  \institution{University of Michigan}
  \city{Ann Arbor}
  \state{Michigan}
  \country{United States}
}
\email{rosianan@umich.edu}

\author{Marquise D Singleterry}
\affiliation{%
  \institution{University of Michigan}
  \city{Ann Arbor}
  \state{Michigan}
  \country{United States}
}
\email{singlete@med.umich.edu}

\author{Filippos Bellos}
\affiliation{%
  \institution{University of Michigan}
  \city{Ann Arbor}
  \state{Michigan}
  \country{United States}
}
\email{fbellos@umich.edu}

\author{Brian George}
\affiliation{%
  \institution{University of Michigan}
  \city{Ann Arbor}
  \state{Michigan}
  \country{United States}
}
\email{bcgeorge@med.umich.edu}

\author{Gurjit Sandhu}
\affiliation{%
  \institution{University of Michigan}
  \city{Ann Arbor}
  \state{Michigan}
  \country{United States}
}
\email{gurjit@med.umich.edu}

\author{Jason J Corso}
\affiliation{%
  \institution{University of Michigan}
  \city{Ann Arbor}
  \state{Michigan}
  \country{United States}
}
\email{jjcorso@umich.edu}

\author{Anhong Guo}
\affiliation{%
  \institution{University of Michigan}
  \city{Ann Arbor}
  \state{Michigan}
  \country{United States}
}
\email{anhong@umich.edu}

\author{Vitaliy Popov}
\affiliation{%
  \institution{University of Michigan}
  \city{Ann Arbor}
  \state{Michigan}
  \country{United States}
}
\email{vipopov@umich.edu}

\author{Xu Wang}
\affiliation{%
  \institution{University of Michigan}
  \city{Ann Arbor}
  \state{Michigan}
  \country{United States}
}
\email{xwanghci@umich.edu}

\renewcommand{\shortauthors}{Wang et al.}

\begin{abstract}
Surgical videos are a primary resource for teaching trainees anatomy, tool
usage, and procedural skills. Yet learning from
them at scale requires systems that understand surgical scenes. Existing
approaches fall short: vision--language models lack fine-grained domain
reasoning, task-specific models do not generalize, and prior scene graphs
omit clinically meaningful detail. We present SurgGraph, a training-free
pipeline that generates quantitative scene graphs from surgical videos.
Operating on segmentation masks and depth maps, SurgGraph encodes each
clinically meaningful relation (attachment, occlusion, separation, tool
actions) as a $\langle$subject, verb, object, value$\rangle$ tuple whose
numeric value quantifies the relation's extent over time. Technical
evaluations show more precise scene understanding than state-of-the-art
surgical VLM baselines. We then build SurgGraphQA, a proof-of-concept
learning application that retrieves meaningful and boundary-case exemplars
and generates visual explanations and feedback. A study with 17 medical
students and 2 resident surgeons shows significant learning gains,
demonstrating its educational value.

\end{abstract}

\begin{teaserfigure}
  \includegraphics[width=\textwidth]{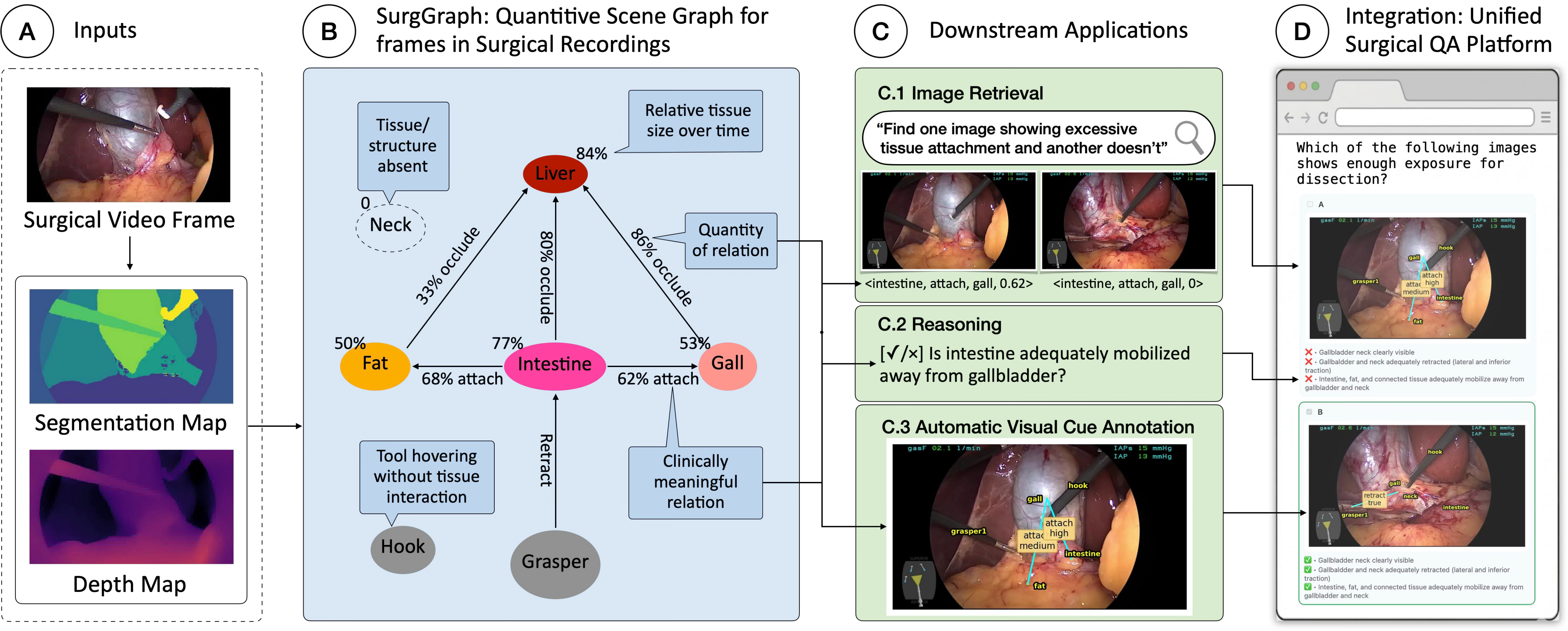}
  \caption{From (A) a frame in the surgical video with its segmentation and depth maps, SurgGraph generates (B) a quantitative scene graph encoding clinically meaningful relations (attachment, separation), their extent, and tool\textendash tissue interactions. This representation enables three applications: (C.1) retrieval of frames matching target scene compositions, (C.2) reasoning on surgical questions, and (C.3) automatic labeling of visual features\textemdash integrated into (D) SurgGraphQA, a proof-of-concept learning application in which trainees learn visual cues for surgical decision making.}
  \Description{}
  \label{fig:teaser}
\end{teaserfigure}

\maketitle

\section{Introduction}
Surgical video recordings are a rich source of medical knowledge \cite{kim2023surch,wang2024surgment, Singh2015, esposito2022video, mazer2022video}. They capture the standard workflows \cite{King2025Use, Loukas2018Video} of clinical procedures, demonstrate expert technique on real patients \cite{Youssef2023Evolution}, and document the nuanced decision-making that defines surgical mastery \cite{Alghazawi2024Development}. Making these recordings computationally understandable unlocks a new generation of intelligent educational tools: systems that answer trainees' questions about specific moments in a procedure or automatically generate explanations of surgical steps.

These applications require a systematic understanding of which entities are present in a video scene, how they relate to each other, and to what degree, since surgery is fundamentally a discipline of controlling relationships between tissues, and between tools and tissues \cite{Trute2024Visual, Shinde2024JIGGLE}. However, existing approaches fall short in three ways. First, general-purpose VLMs \cite{openai2024gpt4o, google2024gemini, bai2023qwenvl} lack \emph{surgical domain knowledge}: benchmarks show that performance degrades substantially on tasks requiring surgical vocabulary and fine-grained visual discrimination \cite{May25,Rau25,Liu25}. Surgery-specific VLMs such as SurgVLM \cite{Zen25} improve domain coverage but still achieve only 63.46\% on surgical VQA \cite{Kha25b,Che25b}. Second, all VLMs lack \emph{numerically grounded understanding}: their outputs are natural language descriptions generated from learned embeddings, which cannot measure the \emph{extent} of a relation but only describe it in approximate language. This means they cannot reliably distinguish between partial and complete tissue detachment within a single frame, nor track how relations change quantitatively across frames, such as the progressive detachment of fat from the gallbladder over the course of a dissection.

These gaps point to the need for a different representation: \emph{quantifiable scene graphs}, in which every relation carries a continuous numeric value capturing not just \emph{whether} two entities are related but also \emph{the extent} of that relationship. 
This representation will enable
(1) retrieval of surgery images using specific relation values that reflect precise visual cues for decision making, e.g., when exposure is sufficient for dissection;
(2) richer reasoning about surgical procedures, e.g., by representing the spatial relationships between anatomical structures and tools; 
and (3) precise annotations of visual cues, including both the relationship and its magnitude.

How can such quantifiable scene graphs be generated without the large annotated datasets that the surgical domain lacks? Prior work has shown that neuro-symbolic approaches that apply programs, such as geometric methods, on neural outputs can extract visual information without training~\cite{zhang2021screen, barnaby2024photoscout, herskovitz2024programally, armeni20193d}. We extend this beyond entity-level extraction to compute quantitative \emph{relations}: neural foundation models provide perceptual inputs (segmentation, depth), and geometric rules grounded in surgical knowledge compute continuous relational values between anatomical structures and instruments.

We present \textbf{SurgGraph}, a training-free pipeline that generates quantifiable scene graphs from surgery recordings. Given segmentation masks from SAM 3~\cite{carion2025sam} and depth maps from Video Depth Anything~\cite{chen2025video}, SurgGraph computes per-frame scene graphs in which each edge is a $\langle$subject, verb, object, quantity$\rangle$ tuple, where the quantity is derived from deterministic geometric operations: attachment is measured as shared boundary length (e.g., $\langle$fat, attached, gallbladder, 0.73$\rangle$), separation as the gap area between structures, and tool activity as depth-verified proximity between an instrument tip and its target. Because every relation is a continuous value tracked across frames, the resulting representation supports reasoning about surgical progress over time.

In particular, we apply SurgGraph to lap chole surgery recordings (also known as gallbladder removal surgeries, the most common minimally invasive surgery). 
Technical evaluations on established benchmarks~\cite{nwoye2023cholectriplet2021, endoscope2023cvs} show that SurgGraph achieves 79.2\% on tool-action-triplet recognition, 93.8\% on phase recognition, and 94.59\% on assessment of the critical view of safety (a key visual decision-making task in lap chole). In comparison, SurgVLM~\cite{Zen25}, a state-of-the-art surgery-specific VLM, achieves 31.3\%, 91.0\%, and 79.4\% on the same tasks respectively.

We further demonstrate the utility of SurgGraph through SurgGraphQA, a
proof-of-concept educational application that leverages the quantitative
scene graphs to generate visual exercises helping medical trainees learn the
visual cues for surgical decision-making.
SurgGraphQA demonstrates three key functionalities supported by the quantitative scene graphs, which are not feasible with VLM-based approaches alone, including 1) \emph{retrieving surgery images that are clinically meaningful and pedagogically interesting}, such as scenes where ``fat attachment to the gallbladder is high and the grasper is retracting the fat'' or diverse images sampled across the progression of the procedure to illustrate 
how the scene evolves; (2) \emph{providing in-depth, education-aligned explanations of surgical scenes}, e.g.,  whether two anatomical structures (cystic duct and artery) are sufficiently separated; and (3) \emph{annotating visual cues}, including attachment regions, separation gaps, and tool-tissue contact. 

A study with 17 medical students shows that participants had significant gains on surgical decision-making knowledge and skills after learning with SurgGraphQA. The average performance increases from 2.41 to 8.29 (out of 10 points) on exposure identification and from 3.12 to 9.53 on CVS assessment respectively. 
SurgGraphQA enhances learning by sampling diverse and clinically meaningful images across procedural progression to help trainees recognize critical landmarks, the magnitude of detachment and separation relationships, tool handling techniques, and decision-making criteria. 

Two resident surgeons, serving as expert reviewers, rated the generated exercises as clinically accurate and of high quality.

\vspace{.3pc}\noindent\textbf{Contributions:}
\begin{itemize}[leftmargin=*,noitemsep,topsep=0pt]
\item \textbf{SurgGraph}, a training-free pipeline for generating quantitative, geometry-grounded scene graphs from surgical videos using a $\langle$subject, verb, object, quantity$\rangle$ representation computed from segmentation and depth maps, with technical evaluation demonstrating higher-precision scene representations on both established benchmarks and novel quantitative relations.
\item \textbf{SurgGraphQA}, a proof-of-concept educational application built on the
quantitative scene graphs to generate visual exercises. A study with 17
medical students shows significant learning gains on surgical decision-making
tests, demonstrating the value of SurgGraph for downstream applications that
require precise video understanding.
\end{itemize}
\vspace{.3pc}

More broadly, SurgGraph demonstrates how neuro-symbolic approaches can advance video understanding in ways that matter to the HCI community: by producing representations that are not only scalable, but also structured and actionable within interaction systems. In high-stakes domains such as surgery recording understanding and surgical education, users need system outputs that support precise retrieval, explanation, and annotation rather than black-box predictions alone. SurgGraph shows that combining foundation-model outputs with explicit geometric reasoning can enable video understanding pipelines that more reliably support user goals, such as learning the visual cues that underlie surgical decision-making. 

\section{Related Works}
\subsection{Surgical Video Understanding}
Surgical video understanding encompasses a hierarchy of tasks of increasing complexity \cite{li2024surgvlm}. At the foundational level, models perform visual perception tasks such as instrument recognition and localization, tissue recognition, and tissue localization \cite{yanik2022deep, ismail2019accurate, castro2019towards}. Building on these extracted visual semantics, more complex tasks requiring temporal analysis become possible, including surgical phase recognition \cite{garrow2021machine, shi2022attention, loukas2018surgical, twinanda2016endonet}, action recognition \cite{yen2025automated, nwoye2023cholectriplet2021, li2023mt}, and workflow analysis \cite{li2024deep, hashemi2025video, li2025surgical}. At the highest level, these cues serve as context for fine-grained reasoning tasks such as skill assessment \cite{dick2024, mcqueen2019video, pugh2021and, feldman2020sages}, Critical View of Safety assessment (method of identification of the cystic duct and cystic artery to prevent bile duct injuries during laparoscopic cholecystectomy) \cite{Vettoretto2011CriticalVO, endoscope2023cvs}, and error detection \cite{ghamrawi2025rectal, shao2024think, gao2014jhu, li2022runtime}. 

More recently, general-purpose VLMs \cite{Sch24, Yao24} have been explored for surgical video understanding, but they are trained predominantly on natural images and text, often generating verbose, clinically irrelevant outputs with ambiguous interpretations that undermine their reliability in surgical practice \cite{May25, Rau25, Liu25, Kha25b, Che25b}. This has motivated efforts toward specialized surgical foundation models. Among these, SurgVLM \cite{Zen25} represents the most advanced surgery-specific VLM to date, integrating visual and language modalities for tasks such as tool-action-triplet identification, surgical VQA, and phase recognition. However, its performance remains insufficient for practical clinical use (e.g., 63.46\% on surgical VQA benchmarks reported by the authors). Moreover, it still follows a data-driven paradigm fundamentally limited by the scarcity of large-scale annotated surgical datasets. We compare SurgGraph against SurgVLM in our technical evaluation (Section~\ref{sec:eval}).

\subsection{Scene Graph Generation}
Scene graph generation (SGG) structures visual scenes into graphs representing entities and their relationships~\cite{xu2024llava,guo2024tri,kim2024scene,chen2024scene,zhang2024scenellm,kim2024llm4sgg,yu2023visually}. Unlike flat classification outputs, scene graphs explicitly model entities and capture their spatial, temporal, and functional relationships. In the surgical domain, CholecTriplet~\cite{nwoye2023cholectriplet2021, nwoye2022rendezvous} introduced action triplet recognition, predicting \textit{(instrument, verb, target)} combinations from laparoscopic video. SurgVQA~\cite{Yuan2023Advancing} generates spatial scene graphs with spatial relations (e.g., ``above,'' ``left of'') for surgical question answering, such as ``Where is bipolar forceps located?,left-top.'' However, existing methods capture only tool actions (e.g., ``tool applied to tissue''), which are of limited accuracy, and spatial relations (e.g., ``tissue A above tissue B'') that convey obvious positional information rather than clinically meaningful measurements to support educational or decision-support applications.

\subsection{Representations for Visual Content}
Constructing structured representations from raw visual input to enable downstream applications has been broadly explored in HCI for videos and UIs. For videos, prior work has extracted step-by-step structures from how-to videos for navigation and summarization~\cite{kim2014toolscape, truong2021automatic}, aligned multi-level textual descriptions to movie timelines for search and browsing~\cite{pavel2015sceneskim}, and defined sentence-level taxonomies over instructional content~\cite{yang2023beyond}. These representations capture the temporal flow of a video but not the visual details within each frame.

For UIs, structured representations more closely resemble what surgical frames need: each frame contains discrete elements with spatial relationships. Rico~\cite{deka2017rico} and Screen2Vec~\cite{li2021screen2vec} learn structured embeddings from UI screens through data-driven models, but require substantial labeled datasets that are far harder to obtain in the surgical domain. Screen Recognition~\cite{zhang2021screen} is closer to SurgGraph: it applies neural object detection to identify UI elements, then uses geometric heuristics over bounding boxes (containment, alignment, overlap) to construct structured accessibility metadata. This follows a neuro-symbolic pattern,neural perception followed by rule-based geometric reasoning,similar to broader neuro-symbolic approaches that decompose visual reasoning into neural perception and symbolic programs~\cite{suris2023vipergpt, gupta2023visual, Mao19, Burghouts24}, a paradigm also adopted in HCI for data-efficient visual systems~\cite{herskovitz2024programally, barnaby2024photoscout}. However, Screen Recognition uses geometry only to classify individual widget properties (is-a-group, is-a-list), not to compute relations between elements. SurgGraph adopts the same principle of separating perception from reasoning, but extends it in two ways: computing relations \emph{between} entities, and preserving the computed geometric values as a measure of relation extent.

\subsection{Geometric Visual Understanding}
Geometric methods in computer vision range from low-level primitives such as edge detection~\cite{canny2009computational}, feature descriptors~\cite{lowe2004distinctive}, and active contours~\cite{kass1988snakes} to higher-level structured outputs. 
In HCI, Prefab~\cite{dixon2010prefab} matches pixel-level widget parts (corners, edges, fills) against geometric templates to recover UI hierarchies from screenshots. In the surgical domain, geometric constraints such as trocar insertion points have been used for line-based instrument localization~\cite{voros2007automatic}. Both use geometry to identify individual entities (widgets, instruments) rather than to compute relationships between them.

More recent work has applied geometry to relational reasoning. 3D Scene Graph~\cite{armeni20193d} computes containment and adjacency between objects using geometric rules over point clouds. However, the computed geometric values (distances, overlap volumes) are thresholded into categorical labels (``adjacent,'' ``inside'') and discarded.

SurgGraph makes a different design choice: the computed geometric values (e.g., attachment = shared edge length) \emph{are} the representation, preserved as scene graph edge weights rather than collapsed into categorical labels. SurgGraph studies how these continuous values can be utilized for more nuanced surgical reasoning.

\section{Context: Laparoscopic Cholecystectomy}
We contextualize our work in laparoscopic cholecystectomy (lap chole), also commonly known as the gallbladder removal surgery, which is the most common minimally invasive procedure.
The American Board of Surgery assessment guidelines~\cite{abs2024assessments}, together with prior research \cite{wang2024surgment}, suggests four main learning objectives for lap chole: anatomy recognition, tool usage, assessment of adequate exposure, and assessment of the critical view of safety. 

In this section, we explain medical terms related to the four learning objectives (LOs) that every general surgery resident is expected to master. These LOs are fundamentally relational, involving the spatial and functional relationships between anatomical structures and surgical tools. This observation motivates the quantitative scene graph design using "geometry-grounded" reasoning. 
\textit{SurgGraph is therefore designed to computationally infer the relationships that underlie these learning objectives.} In this paper, we demonstrate SurgGraph based on this one surgery case, and we will discuss its generalizability to other surgical domains in the Discussion section. 

\emph{LO1: Anatomy recognition.} Identifying anatomical structures in a surgical image requires knowing which structures are present. This is a prerequisite for all relational reasoning.

\emph{LO2: Tool usage.} Learning tool usage involves understanding which instrument should be applied to which structure under what conditions, corresponding to \emph{tool-tissue relations}. Each surgical tool has specific functions and a corresponding set of targets. 

Table~\ref{tab:tool-action-target} summarizes the tool-action-target triplets relevant to lap chole.

\emph{LO3: Assessment of adequate exposure.} Adequate exposure is a prerequisite for safe dissection. According to ABS surgical guidelines~\cite{abs2024assessments}, it requires: (1) Calot's triangle (covered by gallbladder neck) is cleared of all fatty and fibrous tissue; (2) the cystic pedicle (covered by gallbladder neck) is visible; and (3) the gallbladder is properly retracted. Computationally, assessing exposure requires detecting three relations: i) \emph{attachment} between tissues (fat, intestine, connective tissue) and the gallbladder or gallbladder neck is low; ii) \emph{occlusion} of the gallbladder and neck by surrounding tissue is low; and iii) \emph{tool actions} (grasper retraction) is present.

\emph{LO4: Assessment of the critical view of safety (CVS).} The critical view of safety is a widely adopted safety checkpoint before clipping and dividing the cystic duct and artery~\cite{Vettoretto2011CriticalVO}. Achieving CVS means: (1) the Calot's triangle, which is the area enclosed by cystic duct and cystic artery, is cleared of fat and fibrous tissue; (2) the cystic duct and cystic artery are clearly visible as skeletonized tubular structures, well separated from each other; and (3) the lower third of the gallbladder is dissected off the liver bed. Computationally, this requires detecting three relations: i) \emph{separation} between the cystic duct and cystic artery is high; ii) \emph{attachment} relations between connective tissue and the cystic duct or artery is low; and iii) \emph{attachment} between the gallbladder and the cystic plate on the liver bed is low.

\emph{Summary of important relations underlying the LOs.}To computationally represent the relations underlying these LOs, the quantitative scene graphs should capture four core relation types and their magnitude: (1) \textbf{attachment} between tissues, (2) \textbf{occlusion} of one tissue by another, (3) \textbf{separation} between tissues, and (4) \textbf{tool actions} on tissues. 

\section{SurgGraph Generation}

SurgGraph computes quantitative scene graphs through a two-stage pipeline: neural foundation models produce per-frame perceptual inputs, and deterministic geometric algorithms operate over these inputs to compute a continuous relation value for each entity pair. No training data is required beyond the foundation models.

\subsection{Relation Algorithms}
The perceptual inputs are segmentation masks from SAM3~\cite{kirillov2023segment}, identifying anatomical structures and tools across 19 categories (see Appendix~\ref{appendix:segments}), and monocular depth maps from Video Depth Anything~\cite{chen2025video}. Below we demonstrate how we design geometric algorithms for four relation types in lap chole; all parameters are summarized in Table~\ref{tab:parameters}.

\subsubsection{Attachment and Occlusion.}
\begin{figure*}[t!]
  \centering
  \includegraphics[width=\textwidth]{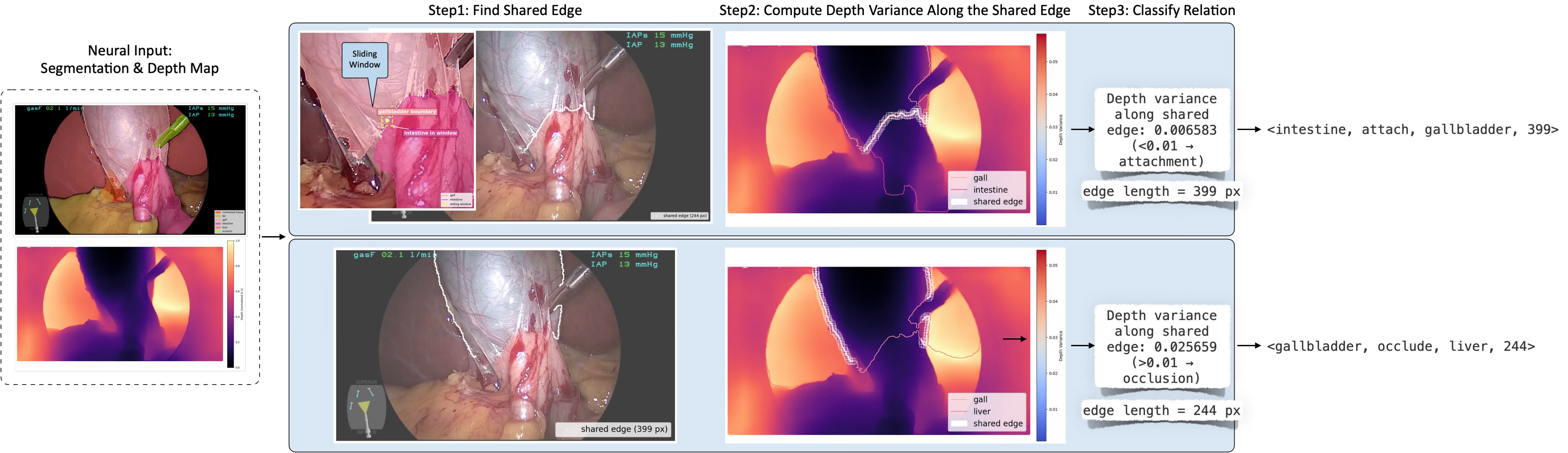}
  \caption{Attachment and occlusion detection.}
  \Description{}
  \label{fig:attachment_occlusion}
\end{figure*}

Attachment and occlusion both involve two tissues whose segmentation boundaries are in close proximity, but they differ in depth continuity along the shared edge. SurgGraph distinguishes them through the following steps (Figure~\ref{fig:attachment_occlusion}):

\noindent \textbf{(1) Find shared edges.} Given the segmentation map, apply a sliding window of size $w$ along the boundary of each tissue segment to check for the presence of a neighboring segment. If a neighbor is found within a proximity threshold, the overlapping boundary pixels form a shared edge between the two tissues.

\noindent \textbf{(2) Compute depth variance along the shared edge.} Using the depth map, slide a window of size $w$ along the identified shared edge and compute the average depth variance across all windows.

\noindent \textbf{(3) Classify the relation.} If the average depth variance is below a threshold $\tau$, the two tissues are at similar depth along their boundary, indicating physical contact: \textbf{attachment}. The shared edge length (in pixels) is stored as the attachment value. If the variance exceeds $\tau$, there is a depth discontinuity along the edge, indicating that one tissue is overlapping the other in the camera view rather than physically touching: \textbf{occlusion}. The shared edge length is stored as the occlusion value.

We illustrate both cases from the same frame in Figure~\ref{fig:attachment_occlusion}. The gallbladder--intestine pair has a shared edge of 244 pixels with low depth variance, indicating attachment. The gallbladder--liver pair has a shared edge of 399 pixels with high depth variance, indicating occlusion.

\subsubsection{Separation.}\label{sec:separation}
\begin{figure*}
  \centering
  \includegraphics[width=0.75\textwidth]{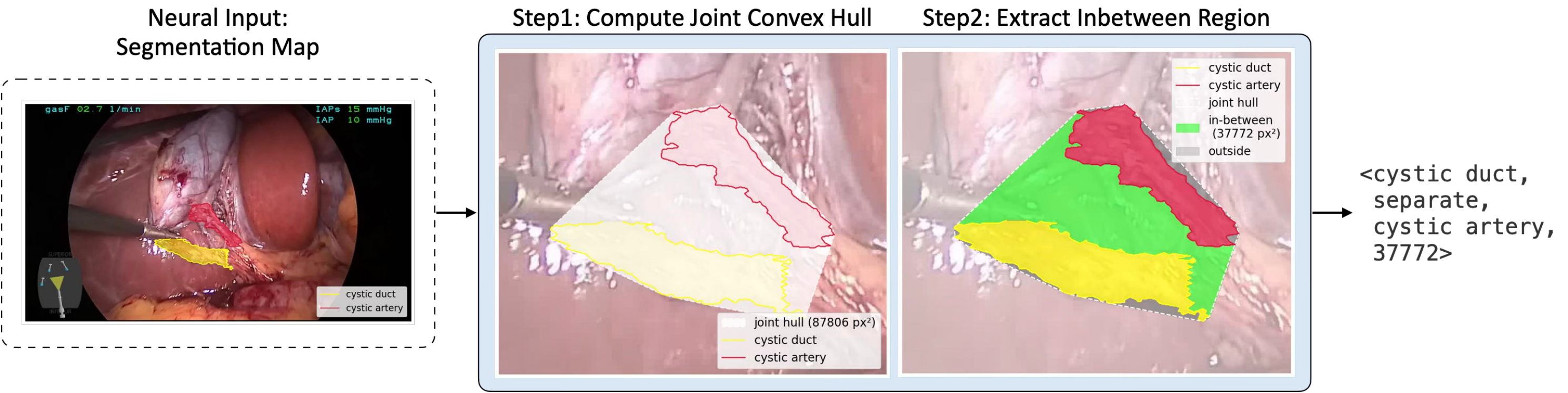}
  \caption{Separation computation between cystic duct and cystic artery.}
  \Description{}
  \label{fig:separation}
\end{figure*}

Separation measures the spatial gap between two structures, which is clinically important for assessing whether the cystic duct and cystic artery have been adequately skeletonized (Figure~\ref{fig:separation}). SurgGraph computes separation in two steps:

\noindent \textbf{(1) Compute the joint convex hull.} Combine the contour points of both structures (e.g., cystic duct and cystic artery) and compute their convex hull. This hull represents the smallest convex region enclosing both structures.

\noindent \textbf{(2) Subtract both segments and extract the in-between region.} Subtract the pixel areas of both segments from the joint convex hull. The remaining space is split into connected components. Components adjacent to only one structure (i.e., outer edge regions of the hull) are discarded. Only the component(s) adjacent to \emph{both} structures are retained, as these represent the actual gap between the two tubular structures. The pixel area of this in-between region is stored as the separation value. A larger value indicates greater spatial separation.

\subsubsection{Tool Actions.}\label{sec:tool-actions}
\begin{figure*}
  \centering
  \includegraphics[width=\textwidth]{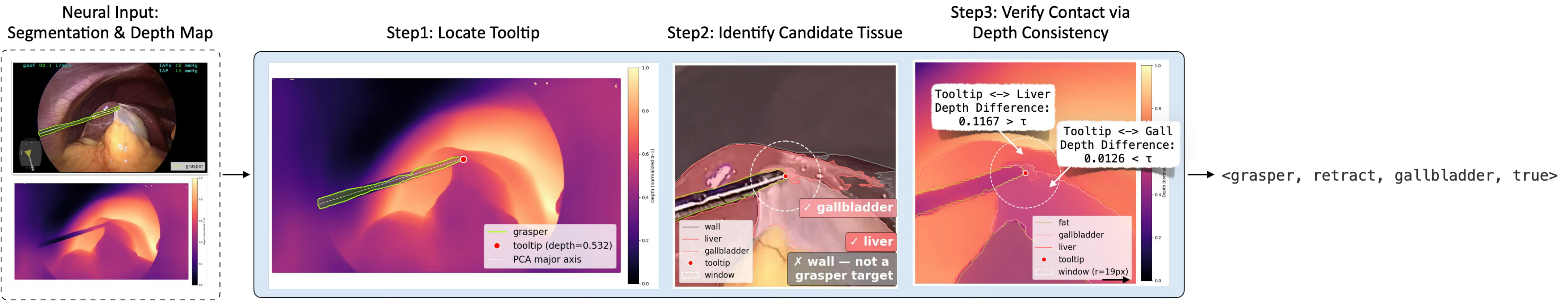}
  \caption{Tool action detection for a grasper.}
  \Description{}
  \label{fig:tool_action}
\end{figure*}

Tool action detection determines whether a surgical instrument is actively contacting a target tissue, producing $\langle$tool, action, target$\rangle$ triplets (Figure~\ref{fig:tool_action}). Each tool type has a fixed set of valid action-target pairs (see Table~\ref{tab:tool-action-target}). The detection proceeds in three steps:

\noindent \textbf{(1) Identify the tool segment and locate the tooltip.} Given the segmentation mask of a tool and the depth map, we locate the tooltip as the deepest point on the tool segment. In laparoscopic surgery, instruments are inserted through trocar ports on the abdominal wall and extend inward toward the surgical site, so the functional tip is always the end furthest from the camera (i.e., the deepest point).

\noindent \textbf{(2) Place a window around the tooltip and identify candidate tissues.} A circular window of radius $r$ is placed around the tooltip. We check which valid target tissues (as defined by the tool type in Table~\ref{tab:tool-action-target}) have pixels falling within this window. Tissues that are not valid targets for the given tool type (e.g., abdominal wall for a grasper) are discarded. If no valid target tissue is present in the window, the tool is classified as having no active contact.

\noindent \textbf{(3) Verify contact via depth consistency.} A tissue being spatially nearby in the image plane does not guarantee physical contact: the tool may be hovering above or below the tissue. To verify contact, we compare the average depth of each candidate tissue within the window to the tooltip depth. If the depth difference exceeds a threshold $\delta$, the tissue is rejected. Among remaining candidates, the tissue with the highest overlap (most pixels in the window) is selected as the contact target, and the corresponding action from Table~\ref{tab:tool-action-target} is assigned to produce the final $\langle$tool, action, target, true$\rangle$ tuple.

\subsection{Parameter Selection}\label{sec:parameters}

\begin{figure*}[t]
  \centering
  \includegraphics[width=\textwidth]{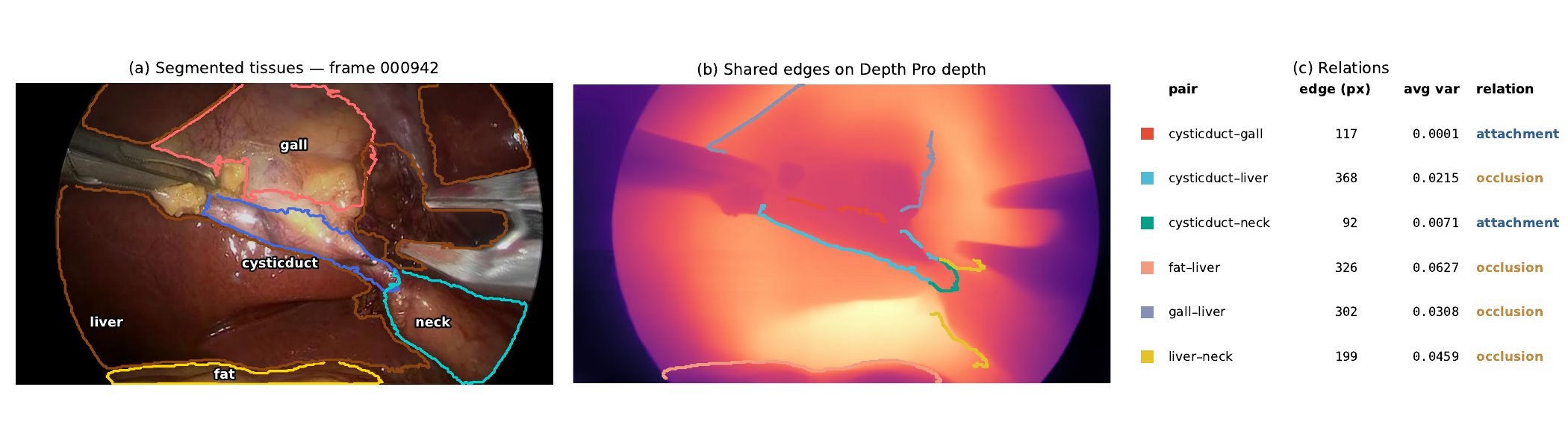}
  \caption{Selecting the depth-variance threshold $\tau$. For the tissue pairs
  in this frame, attachment pairs (physically continuous boundaries) produce
  average depth variances of 0.0001--0.0071, while occlusion pairs (depth
  discontinuities) produce 0.0215--0.0627; $\tau=0.01$ is chosen in the gap
  between the two classes.}
  \Description{A surgical frame with segmented tissues, its depth map with shared edges, and a relations table listing each pair's depth variance.}
  \label{fig:param-tau}
\end{figure*}

\begin{figure*}[t]
  \centering
  \includegraphics[width=0.78\textwidth]{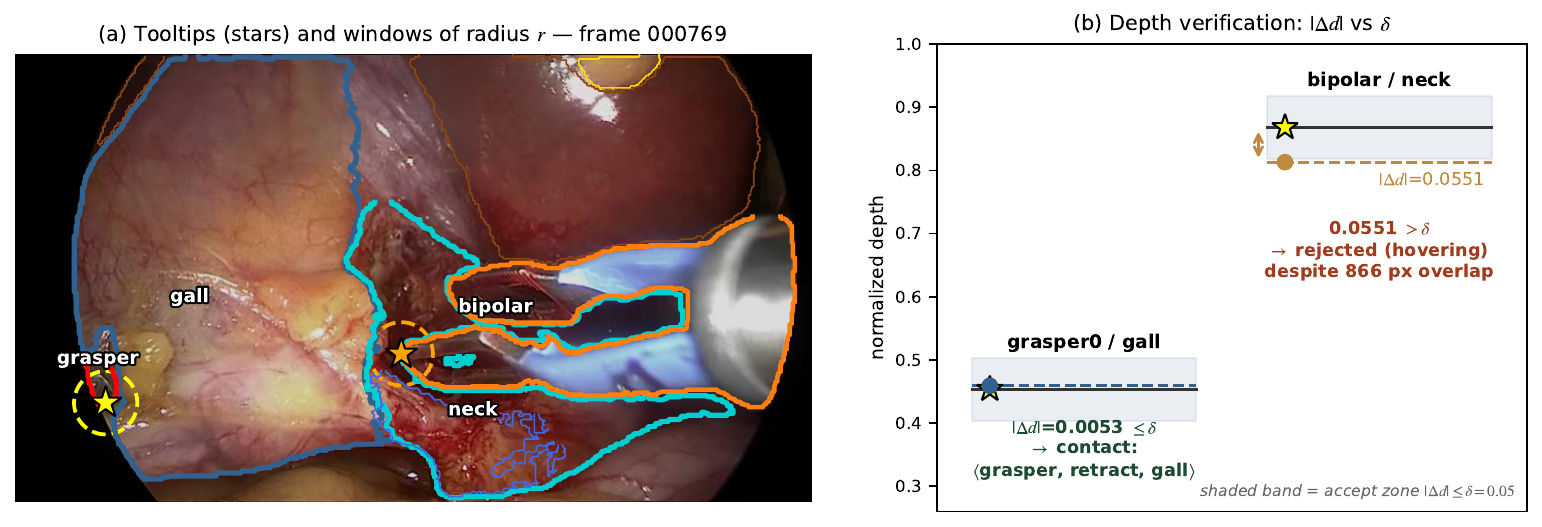}
  \caption{Selecting the depth-difference threshold $\delta$. A contacting
  tool's tip matches the tissue depth closely ($|\Delta d|=0.0053$), while a
  hovering tool differs by an order of magnitude more ($|\Delta d|=0.0551$)
  despite large image-plane overlap; $\delta=0.05$ is chosen between the two.}
  \Description{A surgical frame with two tool windows and a depth-axis diagram showing an accepted contact and a rejected hovering case.}
  \label{fig:param-delta}
\end{figure*}

SurgGraph exposes four parameters (Table~\ref{tab:parameters}), all defined on
normalized quantities and therefore independent of camera resolution and zoom.
The two spatial parameters are set a priori: $w$ spans the boundary uncertainty
of the segmentation masks, and $r$ approximates the apparent size of an
instrument tip. The two depth thresholds are selected by inspecting where the
two classes they separate actually differ. As Figure~\ref{fig:param-tau}
illustrates, physically attached pairs yield depth variances an order of
magnitude lower than occluding pairs (0.0001--0.0071 vs.\ 0.0215--0.0627 in
this frame); $\tau=0.01$ is placed in the gap between the two ranges.
Analogously for tool contact (Figure~\ref{fig:param-delta}): a contacting tool's
tip--tissue depth difference is far smaller than a hovering tool's (0.0053 vs.\
0.0551), and $\delta=0.05$ is placed between them. Both thresholds are then
fixed across all tasks and videos; Section~\ref{sec:sensitivity} shows accuracy
is stable over a wide range around each chosen value. For recordings from a
different camera system, the same selection procedure is repeated on 5--10
expert-labeled frames (Appendix~\ref{appendix:handbook}).

\begin{table}[b!]
\centering
\footnotesize
\begin{tabular}{lll}
\toprule
\textbf{Parameter} & \textbf{Value} & \textbf{Used in} \\
\midrule
Sliding window size $w$ & 2\% of frame width & Attach./Occl. \\
Depth variance threshold $\tau$ & 0.01 & Attach./Occl. \\
Tooltip window radius $r$ & 1\% of frame width & Tool Actions \\
Depth difference threshold $\delta$ & 0.05 (normalized) & Tool Actions \\
\bottomrule
\end{tabular}
\caption{Parameters used in SurgGraph geometric programs.}
\label{tab:parameters}
\end{table}

\subsection{Regularization and Data Structure}\label{sec:regularization}
The raw relation values computed per frame are affected by camera zoom: as the camera moves closer, structures appear larger and relation magnitudes (e.g., shared edge length) increase. To compensate for this zoom effect, we normalize each relation value by the pixel area of a reference anatomical entity that has approximately fixed physical volume throughout the procedure. Specifically, for relations involving the gallbladder (e.g., gallbladder--cystic duct, gallbladder--cystic artery), we divide the raw relation value by the gallbladder's pixel area in that frame. For relations between the cystic duct and cystic artery, we divide by the cystic duct's pixel area. 

After per-entity normalization, we further normalize each pairwise relation across the entire video by dividing by its maximum value over all frames. This scales every relation to the range $[0, 1]$, making values comparable across different relation types and across videos with varying imaging conditions.

The resulting scene graph data for each video is stored as a four-dimensional tensor of shape $F \times 5 \times S \times S$, where $F$ is the number of frames, $5$ corresponds to the five relation channels (attachment, occlusion, separation, tool action, and segment area), and $S$ is the number of segment categories. Each entry encodes the normalized relation value between a subject--object pair for a given relation type at a given frame. For example, querying the attachment relation between the cystic duct (subject) and the gallbladder (object) over time corresponds to the slice $[:,\, \texttt{attachment},\, \texttt{cystic\_duct},\, \texttt{gallbladder}]$. This compact tensor representation enables efficient retrieval of any relation trajectory across the full video.

\section{Technical Evaluation}\label{sec:eval}
We evaluate whether SurgGraph's geometric values accurately reflect the physical scene in two ways: (1) measuring the \emph{accuracy of the values themselves}, and (2) testing whether these values \emph{translate to accurate downstream reasoning}.

For value accuracy, SurgGraph produces two types of values. Tool-action-triplet recognition has established benchmarks with ground-truth labels, enabling direct evaluation. Novel quantitative relations (attachment, occlusion, separation) lack direct ground truth, as estimating continuous values by hand is error-prone. For attachment/occlusion, we estimate a lower bound: if segmentation and relation classification are both correct, the value is exact. For separation, the computed gap between cystic duct and cystic artery corresponds to Calot's triangle in surgical practice (see Figure~\ref{fig:separation-iou} in Appendix), which has ground-truth labels in an open benchmark, enabling IoU evaluation.

For downstream reasoning, SurgGraph's values track surgical progress: attachment between fat and gallbladder decreases as dissection progresses, while separation between the cystic duct and artery increases. Accurate phase identification from these values confirms that the scene is correctly represented. CVS assessment further tests whether values are precise enough for safety-critical clinical reasoning. Both tasks have open benchmarks.

SurgVLM~\cite{Zen25}, the most advanced surgery-specific VLM, serves as the baseline on all benchmarked tasks: tool-action-triplet recognition, phase recognition, and CVS assessment.

\subsection{Value Accuracy}
\paragraph{Baseline.} We compare against SurgVLM~\cite{Zen25}, the
state-of-the-art surgery-specific VLM. Because its model weights are not
publicly released, we re-trained a reimplementation on SurgVLM's released
training data using the same backbone family ([Qwen2.5-VL]), and evaluated it
under a protocol identical to SurgGraph's: the same test frames, the same
label vocabulary, and exact-match scoring. All SurgVLM numbers below come from
this re-run rather than from figures cited in the original paper.
\subsubsection{Tool-Action-Triplet Recognition.}
On the CholecT50 test set~\cite{nwoye2023cholectriplet2021} (1129 frames, 5 videos, $\langle$instrument, verb, target$\rangle$ annotations), SurgGraph predicts triplets via tool-tissue contact detection (Section~\ref{sec:tool-actions}), measured as exact match accuracy. SurgGraph achieves 79.2\%, vs.\ 31.3\% for the re-trained SurgVLM, a 2.5$\times$ advantage. SurgVLM attains high precision (0.984) but very low recall (0.273), predicting only the most frequent triplets, while SurgGraph's explicit geometric steps (tooltip localization, proximity filtering, depth verification) yield balanced precision and recall (0.915/0.878).

\subsubsection{Lower-Bound Accuracy of Attachment/Occlusion Values.}
On CholecT50 frames from the Preparation phase, we labeled relations between five tissue pairs (gallbladder--fat, gallbladder--intestine, intestine--fat, neck--fat, neck--gallbladder), totaling 266 images and 1,133 detected pairs, and evaluated SurgGraph's classification (attachment vs.\ occlusion) against expert labels. Because these values are deterministic functions of segmentation and depth, they are exact whenever both inputs are correct. SAM achieves 92\% segmentation accuracy on a lap chole segmentation benchmark~\cite{wang2024surgment}. Relation classification achieves 82.2\%, yielding a lower bound of $\text{acc}_{\text{lower}} = 0.92 \times 0.822 = 75.6\%$. This confirms that attachment and occlusion can be reliably distinguished using depth variance heuristics, and that the computed extent values are accurate under most conditions.

\subsubsection{Separation Value Accuracy via IoU}
On 74 frames from the EndoVis CVS benchmark~\cite{endoscope2023cvs} with ground-truth Calot's triangle labels, we compute IoU between SurgGraph's separation area (Section~\ref{sec:separation}) and the ground-truth annotation. Average IoU is 0.73. In practice, the separation region is a gap where underlying structures (cystic plate, liver) become visible as the cystic duct and artery are skeletonized, making it difficult to segment as a distinct region. The result confirms that the in-between area identified through convex hull subtraction is a reliable geometric proxy for how well two tubular structures are separated. The IoU is not perfect because residual tissue within the gap is included in the computed area when skeletonization is incomplete, but the overall size and evolving trend of the triangle are well captured, enabling reliable tracking of the dissection process.

\subsubsection{Parameter Sensitivity}\label{sec:sensitivity}

\begin{figure*}[t]
  \centering
  \includegraphics[width=0.8\linewidth]{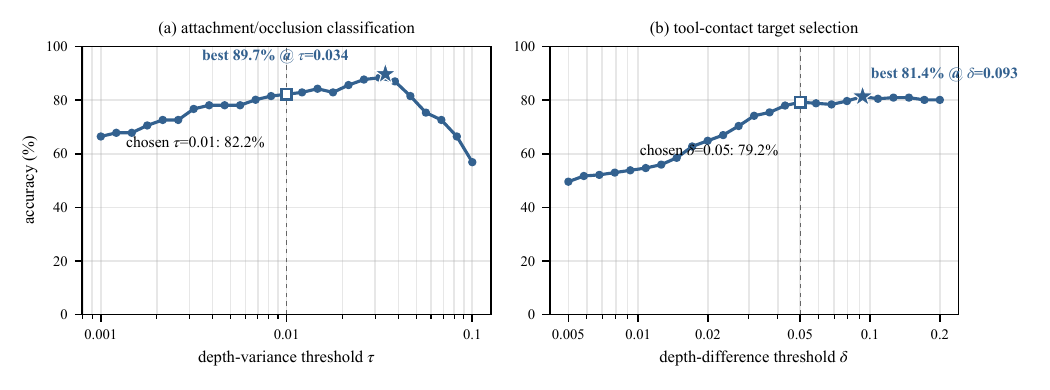}
  \caption{Threshold sensitivity, evaluated against expert annotations. Squares
  mark the chosen operating points ($\tau=0.01$, $\delta=0.05$); stars mark the
  empirical optima. Accuracy is stable over wide ranges around the chosen
  values and degrades only at physically implausible settings.}
  \Description{Two accuracy-versus-threshold curves, each with a broad plateau around the chosen operating point.}
  \label{fig:sensitivity}
\end{figure*}

The two depth thresholds are the only calibrated parameters, so we verify that
SurgGraph's accuracy does not depend on their exact placement. We sweep each
threshold against expert annotations---tissue-pair relation labels for $\tau$
and per-instance tool-contact labels for $\delta$---while holding everything
else fixed (Figure~\ref{fig:sensitivity}).

For $\tau$, accuracy stays within four points of the optimum (89.7\% at
$\tau=0.034$) across the range $[0.02, 0.04]$, and remains above 78\% across a
full order of magnitude $[0.005, 0.05]$; the chosen $\tau=0.01$ yields 82.2\%.
Accuracy degrades only where the threshold becomes physically implausible: far
below the depth model's noise floor, smooth attached boundaries are misread as
occlusions, and far above it, genuine depth discontinuities are absorbed as
attachment. For $\delta$, accuracy rises steeply until the threshold clears the
depth model's error level, then varies by less than 2.2 points across the
4$\times$ range $[0.05, 0.2]$ (best 81.4\% at $\delta=0.093$); the chosen
$\delta=0.05$ sits at the start of this plateau and yields 79.2\%. Setting
$\delta$ below this range rejects true contacts as hovering (49.6\% at
$\delta=0.005$).

\subsection{Downstream Reasoning Accuracy}

\subsubsection{Phase Recognition.}
On the CholecT50 test set~\cite{nwoye2023cholectriplet2021} (1129 frames, 7 phases), we apply rule-based classifiers using conjunctions of scene graph conditions over attachment, separation, tool action, and area values (see Appendix~\ref{appendix:phase-rules}). The classifiers use only simple threshold comparisons with no learned parameters, so accuracy reflects the quality of the underlying geometric values. SurgGraph achieves 93.8\% vs.\ SurgVLM's 91.0\% on the same test set. Accuracy alone understates the difference: macro-F1 is 0.943 vs.\ 0.529, because SurgVLM's accuracy is carried by frequent phases while SurgGraph maintains per-phase F1 between 0.914 and 0.975 across all seven phases.

\subsubsection{Critical View of Safety (CVS) Assessment.}
On the EndoVis CVS benchmark~\cite{endoscope2023cvs} (2,250 frames, 10 videos, frame-level annotations of three CVS criteria), each criterion is mapped to a SurgGraph relation: \textbf{C1} (two structures visible): $\langle$cystic duct, area$\rangle > 60\%$ AND $\langle$cystic artery, area$\rangle > 60\%$; \textbf{C2} (triangle cleared): $\langle$cystic duct, separate, cystic artery$\rangle > 60\%$; \textbf{C3} (lower gallbladder dissected): $\langle$gallbladder, attach, cystic plate$\rangle < 80\%$.

SurgGraph achieves 91.8\% on C1, 90.5\% on C2, 77.0\% on C3, and 94.59\% overall, vs.\ the re-trained SurgVLM's 89.4\%, 81.3\%, 76.5\%, and 79.4\%.. The overall accuracy can exceed individual criterion accuracies because CVS is conjunctive: if any criterion is unmet, CVS is classified as not achieved, so the overall prediction can be correct even when the specific unmet criterion is misidentified. The high C1 and C2 accuracy shows that geometric values are precise enough for assessing structure visibility and the degree of separation. The lower C3 (77.0\%) is because this criterion is inherently difficult to assess: the benchmark uses three raters, and frames without rater agreement are labeled as not met, introducing false negatives. 

\section{SurgGraphQA: A Proof-of-Concept Learning Application}\label{sec:platform}
SurgGraphQA is a proof-of-concept downstream application of SurgGraph, built
to demonstrate how quantitative scene graphs can support education. SurgGraph
is motivated by surgical recordings being widely used as a learning resource,
and SurgGraphQA closes this loop: the videos that trainees would watch are
automatically transformed into structured educational content, so every new
recording can generate new visual exercises, feedback, and labeled examples
without manual authoring. Learning sciences and cognitive science research
has long suggested that practicing with varied examples better supports
knowledge retention and transfer than repeated practice with the same
examples~\cite{butowska2024variable, butler2017transfer, smith1978context},
which motivates presenting medical trainees with diverse operative images
that instantiate the same decision-making principles.

Our goal is not a complete tutoring system. Rather, SurgGraphQA demonstrates
three downstream capabilities that quantitative scene graphs uniquely enable:
retrieving images by their scene composition (Section~\ref{sec:retrieval}),
generating criterion-level explanations of a scene
(Section~\ref{sec:explanation}), and automatically annotating visual cues on
images (Section~\ref{sec:labeling}). Each mechanism below is a direct query
over the $\langle$subject, verb, object, value$\rangle$ tuples, illustrating
what the representation itself makes possible.

SurgGraphQA generates visual exercises from instructor-authored templates,
following evidence that template-based generation produces more reliable items
than free-form automatic generation~\cite{wang2019upgrade, gierl2013aig,
west2015prairielearn}. Beyond SurgGraph's scene graphs, the only additional
input SurgGraphQA requires is the template configuration itself: for each
learning objective, the research team authors (1)~a textual template for the stem and answer options, (2)~scene-graph conditions specifying which frames to retrieve, and
(3)~feedback rules specifying a criteria checklist and which relations or
entities to label on the feedback image. At runtime, SurgGraphQA instantiates
each template by retrieving matching frames, evaluating the criteria against
scene-graph values, and annotating the images accordingly.

The generated visual exercises (both multiple-choice and open-ended) are organized under two procedural topics: \textbf{Exposure} and \textbf{Critical View of Safety}, the two most technically demanding steps in laparoscopic cholecystectomy according to American Board of Surgery resident performance assessments~\cite{abs2024assessments}, each covering anatomy recognition, tool usage, and procedural assessment.
\subsection{Exercise Templates}
Each template specifies a textual template for the stem and answer options,
scene-graph conditions for image retrieval, and feedback rules for
explanation and labeling. Figure~\ref{fig:template-mcq} shows the
procedural-assessment template; the anatomy and tool templates are shown in
Appendix~\ref{appendix:templates} (Figures~\ref{fig:template-anatomy}
and~\ref{fig:template-tool}).

\emph{Anatomy recognition.} (Appendix~\ref{appendix:templates}, Figure~\ref{fig:template-anatomy}). Stem: \emph{``Which structure is [TARGET] in the following image?''} The target is randomly sampled from key landmarks: liver, gallbladder, fat, intestine, cystic pedicle (neck), cystic duct, and cystic artery. Distractors are sampled from the remaining structures. SurgGraph retrieves a frame whose scene graph contains both the target and distractors (queried as $\langle$TARGET, area, 1$\rangle$, $\langle$STRUCTURE\_1, area, 1$\rangle$, etc.), then labels them as A, B, C on the image using segmentation mask locations. Feedback reveals the same image with all structures labeled by their correct names.

\emph{Tool usage.} (Appendix~\ref{appendix:templates}, Figure~\ref{fig:template-tool}). Stem: \emph{``Which tool can be used to [FUNCTION]?''} A tool function is sampled from the tool-action-target dictionary (Table~\ref{tab:tool-action-target}) and converted into a natural language description by GPT-4o~\cite{openai2024gpt4o} (e.g., ``provide traction and retract the gallbladder laterally and inferiorly''). Answer options list candidate tools. Feedback retrieves a frame where the correct tool is performing the described action on its target (queried as $\langle$TOOL, action, TARGET, true$\rangle$), with both the tool and target tissue highlighted.

\emph{Procedural assessment.} (Figure~\ref{fig:template-mcq}). Stem: \emph{``Which of the following images achieves sufficient Exposure / Critical View of Safety?''} SurgGraph retrieves two frames at different procedural stages (Section~\ref{sec:retrieval}), provides a criteria checklist (\checkmark/\texttimes) as textual feedback (Section~\ref{sec:explanation}), and labels the relevant relations and tool actions directly on the images (Section~\ref{sec:labeling}).

\begin{figure*}[t]
  \centering
  \includegraphics[width=0.8\linewidth]{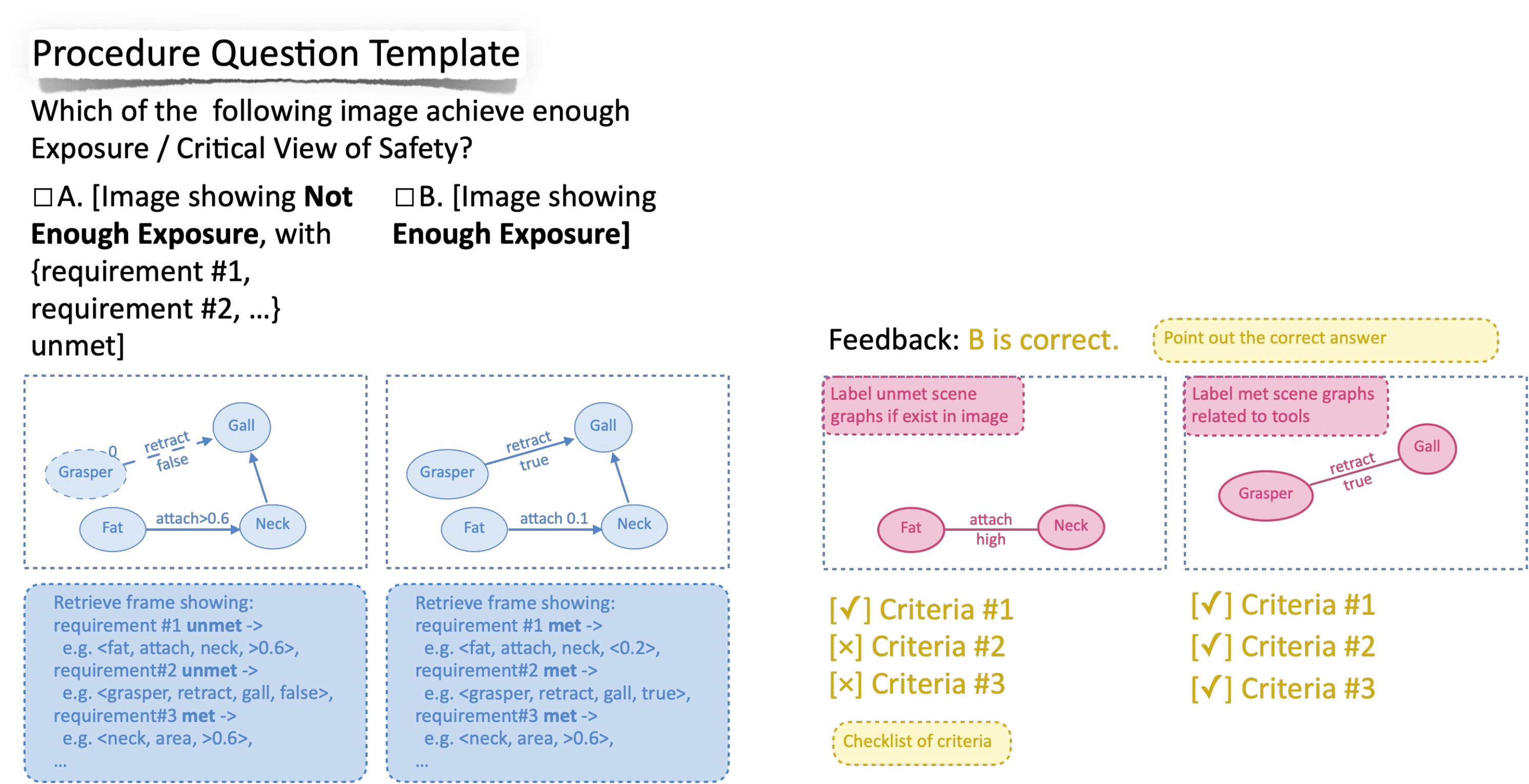}
  \caption{The procedural-assessment template. SurgGraphQA retrieves frames at
  different procedural stages (Section~\ref{sec:retrieval}), presents a
  criteria checklist as feedback (Section~\ref{sec:explanation}), and labels
  the relevant relations and tool actions on the images
  (Section~\ref{sec:labeling}).}
  \Description{An exercise template showing two surgical frames as answer options, a criteria checklist, and labeled feedback images.}
  \label{fig:template-mcq}
\end{figure*}

For open-ended variants, answer options are replaced with a text input box. For procedural open-ended exercises, the criteria checklist is converted into a rubric that specifies which relations and structures must be referenced in the student's response.

\subsection{Utility 1: Image Retrieval by Scene Composition}\label{sec:retrieval}

Image retrieval selects frames from the surgical video that satisfy specific scene graph conditions as required by the exercise templates. We define two retrieval modes:

\begin{figure*}
  \centering
  \includegraphics[width=\textwidth]{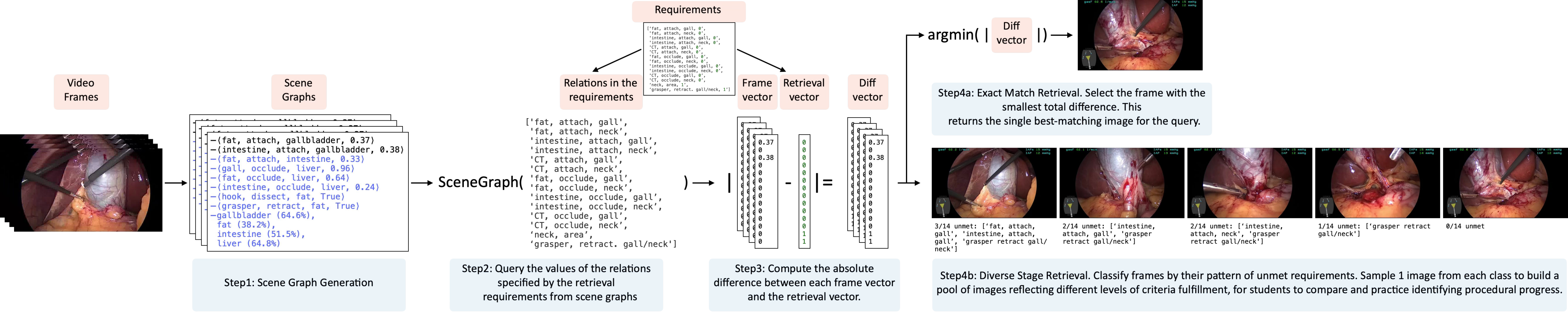}
  \caption{Image retrieval pipeline. Scene graphs are computed per frame, relation values are queried, and the difference from the retrieval vector determines frame selection (Mode A Exact Match Retrieval: best match; Mode B Diverse Stage Retrieval: diverse sampling across unmet requirements).}
  \Description{}
  \label{fig:retrieval}
\end{figure*}

\emph{Mode A: Exact Match Retrieval.} Mode A finds the single frame
that best matches a target scene description. SurgGraphQA encodes the
template's retrieval conditions as a target vector
$\mathbf{r} \in \{0, 1\}^d$, one dimension per condition; for the example
above, $\mathbf{r} = [1, 1]$ (gallbladder present, grasper retracting). Each
frame is likewise encoded as a vector $\mathbf{f}$ of its normalized
scene-graph values. For example, a frame in which the gallbladder occupies
80\% of its maximum visible extent while the grasper is not retracting yields
$\mathbf{f} = [0.8, 0]$. SurgGraphQA then returns the best-matching frame,
$\arg\min_{\mathbf{f}} \|\mathbf{f} - \mathbf{r}\|$. Because the scene graph
stores continuous values for every entity in every frame, this retrieval
targets frames with an optimally clear view of all structures in the
exercise---which embedding-based similarity or keyword matching cannot
guarantee.

\emph{Mode B: Diverse Stage Retrieval.} Procedural queries such as ``sufficient exposure'' are defined by many criteria at once, and some
combinations are mutually exclusive (e.g., the neck cannot be both invisible
and retracted), so no single frame can satisfy them all and direct matching
would miss meaningful surgical states. Mode B instead samples
\emph{across stages of completion}. SurgGraphQA first evaluates every
criterion of the learning objective against each frame's scene graph, treating
a criterion as satisfied when its value is within 0.2 of the target state (the
``largely absent/present'' landmark used throughout,
Section~\ref{sec:regularization}). It then groups frames by their
\emph{pattern of satisfied and unsatisfied criteria}: two visually different
frames fall into the same group if the same criteria remain unmet (e.g.,
``only fat attachment unmet'' vs.\ ``fat attachment and grasper retraction
both unmet''). Finally, it samples representative frames uniformly from each
group (implementation details in Appendix~\ref{appendix:modeb}). Because
relation values evolve with procedural progress, frames from different stages satisfy
different criterion subsets, so the sampled set naturally spans the procedure:
in Figure~\ref{fig:retrieval}, five distinct stages emerge in this video, from
largely unmet (fat heavily attached, no retraction) to nearly complete (only
minor fat residue). Learners therefore encounter a pedagogically meaningful
range of surgical states rather than only fully met or fully unmet examples.

\subsection{Utility 2: Criterion-Level Explanation Generation}\label{sec:explanation}

For anatomy and tool exercises, textual feedback is directly determined by the template: the correct structure name from segmentation labels or the correct tool function from the detected triplet. For procedural exercises, SurgGraphQA provides criterion-level checklist feedback that maps visual cues to specific textbook criteria. These criteria are part of each exercise template's feedback configuration (Section~\ref{sec:platform}): for each learning objective, the research team, together with attending surgeons, operationalizes the textbook requirements for that objective~\cite{abs2024assessments} as compositions of scene-graph conditions over the landmark values of Section~\ref{sec:regularization}. The criteria are thus authored once per learning objective at the application layer; they are not built into SurgGraph itself. For example, the exposure criteria are:
\begin{itemize}[leftmargin=*]
\item $\langle$fat/intestine/CT, attach/occlude, gall/neck, $<$0.2$\rangle$ $\rightarrow$ ``Intestine, fat, and connective tissue adequately mobilized away from gallbladder and neck''
\item $\langle$neck, area, $>$0$\rangle$ $\rightarrow$ ``Gallbladder neck clearly visible''
\item $\langle$grasper, retract, gall/neck, true$\rangle$ $\rightarrow$ ``Gallbladder and neck adequately retracted''
\end{itemize}
The CVS criteria follow the same pattern (see Appendix~\ref{appendix:cvs-criteria}). The feedback presents each criterion as a checklist (\checkmark/\texttimes) based on the computed scene graph values. Because the values are quantitative and grounded in geometric computation, the checklist is more reliable than free-text explanations from a VLM, which cannot guarantee that each criterion is individually and correctly assessed. For open-ended questions, these criteria are converted into a rubric specifying which relations and structures must be referenced in the student's response.

\subsection{Utility 3: Automatic Visual Cue Annotation}\label{sec:labeling}
\begin{figure*}[t]
  \centering
  \includegraphics[width=\textwidth]{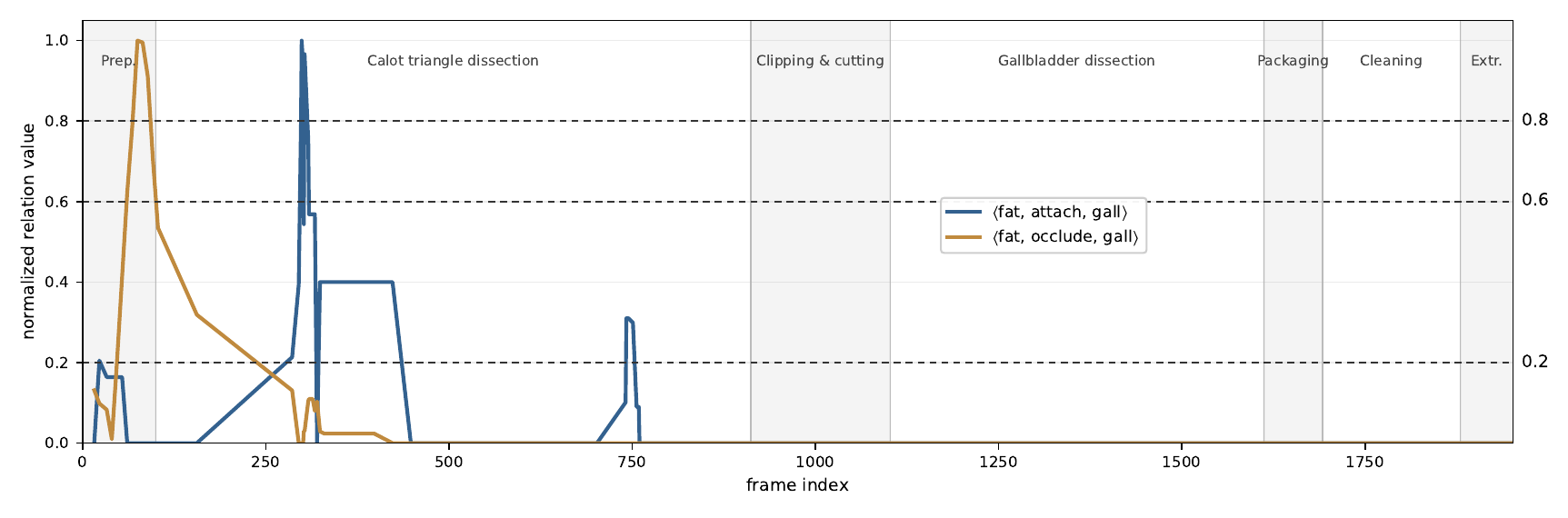}
  \caption{Normalized relation values across a full procedure (smoothed and
  max-normalized; surgical phases shaded), grounding the labeling levels. The
  fat--gallbladder attachment and occlusion values are high while fat still
  covers or adheres to the gallbladder and fall below 0.2 once mobilization
  completes; the cystic duct--artery separation lies predominantly above 0.6
  during clip application, when surgeons have judged the structures clearly
  separated. The level boundaries (0.2, 0.6) thus correspond to clinically
  meaningful transitions rather than tuned parameters.}
  \Description{Three smoothed curves over frame index with surgical phase bands: two fall below the 0.2 line after mobilization; one rises above the 0.6 line during the clipping phase.}
  \label{fig:relation-timeseries}
\end{figure*}

\begin{figure*}
  \centering
  \includegraphics[width=\textwidth]{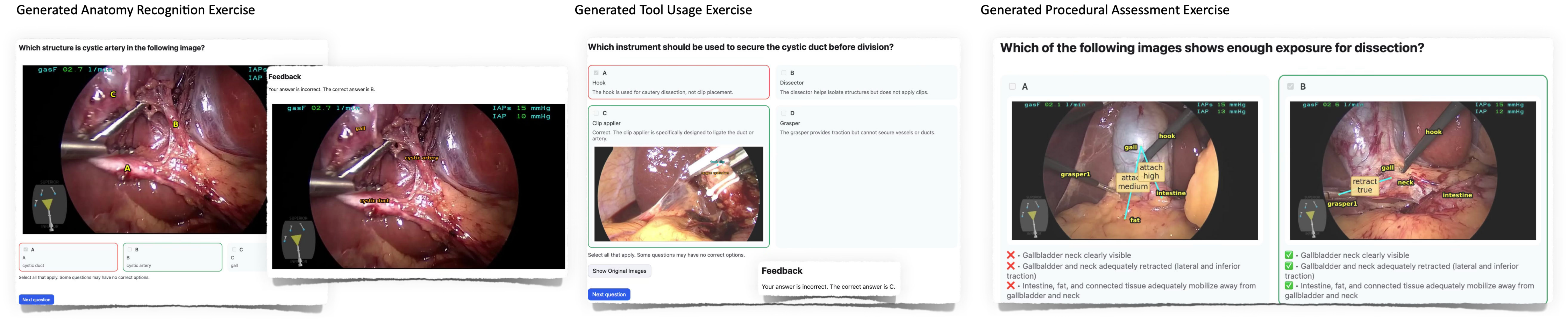}
  \caption{Generated visual exercises for anatomy recognition, tool usage, and procedural assessment.}
  \Description{}
  \label{fig:labeling}
\end{figure*}

To reinforce visual identification, we label relevant features on feedback images (Figure~\ref{fig:labeling}). For anatomy exercises, structure names are labeled at their segmentation mask locations. For tool exercises, the tool and its target tissue are labeled (e.g., ``grasper $\rightarrow$ gallbladder''). For procedural exercises, we additionally label:

\textbf{Attachment:} regions where tissues are attached at medium ($0.2$--$0.6$) or high ($> 0.6$) levels.

\textbf{Tool action:} active tool-tissue contact (e.g., ``grasper retracting gallbladder'').

\textbf{Separation:} degree of separation between cystic duct and cystic artery, labeled as low ($< 0.2$), medium ($0.2$--$0.6$), or high ($> 0.6$).

 Because each scene graph value traces back to specific pixel regions, SurgGraph can annotate exact locations and provide a sense of extent through levels (low, medium, high). 

These level boundaries reuse the same 0.2/0.6 landmarks applied throughout the
paper rather than per-label tuning; Figure~\ref{fig:relation-timeseries}
grounds them in the procedure itself: relation values fall below 0.2 as the
corresponding surgical goal is completed (e.g., fat mobilized away from the
gallbladder), while the duct--artery separation lies predominantly above 0.6
precisely during clip application.

\section{SurgGraphQA User Evaluation}
We conducted a user study to examine whether the exercises SurgGraphQA
generates are pedagogically meaningful and clinically accurate---that is,
whether the fully automatic path from surgical video to visual exercises
(Section~\ref{sec:platform}) yields learning experiences that trainees find
valuable and that experts endorse. As a feasibility evaluation of a novel
exercise-generation pipeline, the study does not include a comparison
condition: our goal is to characterize what trainees learn from and value in
the generated exercises, rather than to isolate the causal contribution of the
scene-graph representation against alternative instruction (see
Section~\ref{sec:limitations}). Specifically, we investigate the following
research questions:
\begin{itemize}[leftmargin=*]
\item \textbf{RQ1:} Does learning with SurgGraph-generated content improve surgical knowledge?
\item \textbf{RQ2:} Do learners find the SurgGraph-enabled features educationally valuable, specifically: (a) anatomy and tool examples retrieved from multiple videos, (b) examples from multiple procedural stages, (c) criteria-linked feedback, and (d) labeled images showing relation extent?
\end{itemize}
\subsection{Study Materials}

We generated visual exercises from four laparoscopic cholecystectomy videos: two publicly available educational videos from YouTube~\cite{youtube2024lapchole1, youtube2024lapchole2} and two full-length ($\sim$2 hour) recordings collected with IRB approval from a U.S. teaching hospital. Exercises were organized into two topics: \textbf{Exposure} and \textbf{Critical View of Safety}. Under each topic, exercises covered anatomy recognition, tool usage, and procedural assessment. The exercise set included both multiple-choice and open-ended formats, with images retrieved and labeled by SurgGraph as described in Section~\ref{sec:retrieval}--\ref{sec:labeling}. In total, SurgGraphQA contained 30 visual exercises across the two topics.

To measure learning gain (RQ1), two attending surgeons created separate pretest and posttest question sets for the two topics (8 questions each per section) based on  videos, adapted to the online question format used in the study. These questions did not overlap with the training exercises in the main learning session, ensuring that learning gain reflects knowledge acquisition rather than memorization.

\subsection{Participants}

We recruited 19 participants (17 medical students, 2 resident surgeons) from a U.S. teaching hospital. Participants had varying levels of surgical experience, ranging from pre-clinical students to senior residents. Full demographic details are provided in Table~\ref{tab:demographics} (Appendix). The study was IRB approved.

\subsection{Study Procedure}

The study followed a pretest--intervention--posttest design and takes approximately 60 minutes per participant. Each participant completed two sections (Exposure and CVS); the order of the two sections was counterbalanced across participants. Each section consisted of:

\begin{enumerate}[leftmargin=*]
\item \textbf{Pretest} (4 question: 1 anatomy+1 tool+2 procedural): Participants answered MCQs and open-ended questions without feedback.
\item \textbf{Main session} (15 visual exercises: 4 anatomy+4 tool+7 procedural assessment): Participants interacted with SurgGraphQA. After submitting each answer, they received textual feedback (criteria checklist with \checkmark/\texttimes) and labeled feedback images. For open-ended exercises, participants could amend their response until correct, at which point the next exercise became available.
\item \textbf{Posttest} (1 anatomy+1 tool+2 procedural): Participants answered MCQs without feedback to measure learning gain. The two sets are isomorphic forms: matched one-to-one in question type, targeted structure or criterion, and difficulty, differing only in the retrieved frames. Form assignment was counterbalanced across participants, so
each form served as the pretest for half of the participants and as the
posttest for the other half.
\end{enumerate}
\subsection{Measures}

\paragraph{Learning gain.} We measured the difference in pretest and posttest scores to quantify knowledge acquisition. Each pretest/posttest section is 10 points, scored as follows: anatomy questions earn 1 point each, tool questions earn 1 point each, and each procedural question earns up to 4 points (1 for the yes/no judgment plus 1 for each of the 3 criteria correctly identified). The rubrics for each question are provided in Appendix~\ref{appendix:rubrics}. One researcher graded all pretest and posttest responses according to the rubrics, blinded to whether each response came from
a pretest or a posttest: responses were pooled and randomly ordered before
grading. Learning gain was computed as raw gain: $\text{learning gain} = \text{posttest} - \text{pretest}$.

\paragraph{Qualitative feedback.} After completing the exercises, participants were interviewed about their experience. The semi-structured interview covered the usefulness of visual labels, image comparisons across procedural stages, diverse patient examples, and textual explanations (criteria checklists and hints).

\subsection{Results}

\subsubsection{Learning gain (N=17 medical students).}
\begin{figure}[t]
  \centering
  \includegraphics[width=0.9\columnwidth]{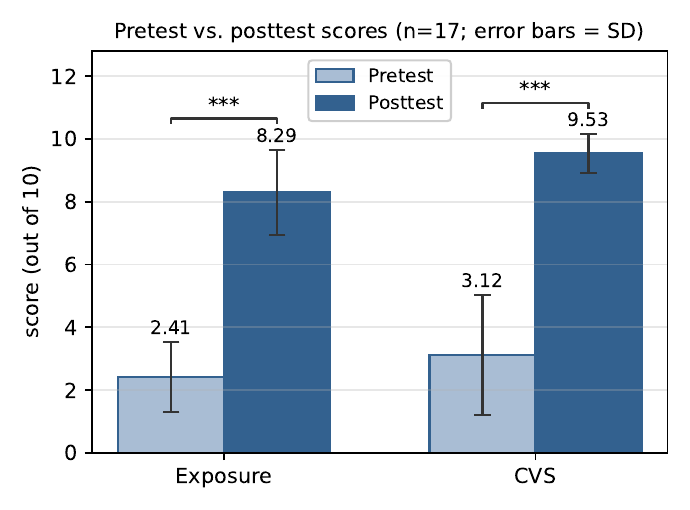}
  \caption{Pretest and posttest scores by section ($n=17$; error bars show
  SD). Both improvements are statistically significant (Exposure:
  $t(16)=14.34$; CVS: $t(16)=13.00$; both $p<.001$).}
  \label{fig:learning-gain}
\end{figure}
Learning gain was measured on the 17 medical student participants; the 2 resident surgeons served as expert reviewers and did not complete the pretest/posttest. Participants showed substantial knowledge gains in both sections (Figure~\ref{fig:learning-gain}).. For exposure knowledge, pretest scores averaged $M=2.41$ ($SD=1.12$) out of 10 and posttest scores averaged $M=8.29$ ($SD=1.36$), yielding a mean raw learning gain of $5.88$ ($SD=1.69$); this improvement was statistically significant ($t(16)=14.34$, $p<.001$, Cohen's $d=3.48$). For CVS knowledge, pretest scores averaged $M=3.12$ ($SD=1.90$) and posttest scores averaged $M=9.53$ ($SD=0.62$), yielding a mean raw learning gain of $6.41$ ($SD=2.03$; $t(16)=13.00$, $p<.001$, Cohen's $d=3.15$). 

\subsubsection{Qualitative findings.} We organized our findings into the following themes from participant interviews (N=19).

\emph{a. Relationship labels with extent help learners understand surgical criteria.}
Labels showing attachment, separation, and spatial relationships helped learners calibrate how much dissection is ``enough.'' P6 said: \emph{``Seeing the labeled images helped me realize the standard for critical view of safety is actually clearer and more skeletonized structures. So it helps you to understand to what extent it should be separated.''} Senior resident P11 and P16 both confirmed the accuracy of the labeling. This validates the core value of quantitative scene graphs for education: learners see not just anatomy but the \emph{relations} and their \emph{extent}, which is what surgical decisions depend on.

\emph{b. Comparing images from different stages and edge cases supports procedural reasoning.}
Seeing images at varying dissection stages moved learning beyond binary safe/unsafe judgments toward identifying which specific criteria are unmet. P1 said: \emph{``Identifying that the critical view of safety is not met is not enough. You have to point out what you need to do as your next step to reach critical view of safety... That would be a deeper level of learning.''} Senior resident P11 found the comparison strategy (perfect vs.\ imperfect CVS) ``really nice'' for pattern recognition. This validates SurgGraph's diverse stage retrieval, which enables comparative learning that is hard to achieve in traditional surgical education.

\emph{c. Automated labeling across diverse videos supports scalable learning.}
Participants emphasized that seeing labeled examples from different patients was essential for generalization. P4 said: \emph{``Every abdomen looks different, so it's important to get exposure to as many of those images as possible.''} Because SurgGraph processes videos automatically, it can scale to large video corpora, making this diversity practically achievables.

\emph{d. Criteria checklists provide structured, objective feedback.}
The checklist format helped learners confirm whether their reasoning was correct. P8 said: \emph{``It was helpful for instances where the CVS was not met, to assess whether the reasons I felt like it was not met were concordant with the reasons that it actually wasn't met.''} Senior resident P11 noted the checklists help learners focus on criteria even with unfamiliar anatomy, as the feedback is grounded in textbook knowledge and consistent across cases: \emph{``At least they have the criteria to go, this is what I'm looking for when I'm making this assessment.''} Senior resident P16 confirmed the feedback was educationally valuable: \emph{``It tells you that it's wrong, but then it tells you why it's wrong.''} This validates SurgGraph's explainable procedural reasoning: criterion-level feedback derived from scene graph values catches reasoning errors that binary correct/incorrect feedback would miss.

\emph{e. Tool usage images pair visual examples with textual explanation.}
Feedback images showing tools in action helped learners understand tool selection. P15 said: \emph{``Because it would say what tools are typically used for, which I think was helpful. I don't know if that's necessarily a primer that we even really got in surgery.''} This validates SurgGraph's ability to retrieve relevant examples of tool--tissue interactions from the video corpus.

\section{Discussion}
\subsection{Why Geometric Rules Still Matter in the Age of VLMs}
A natural question is whether explicit geometric rules are obsolete now that vision-language models (VLMs) can jointly reason over images and text. 

\emph{VLMs struggle with quantitative spatial reasoning.} Recent benchmarks demonstrate that VLMs perform poorly on tasks requiring precise spatial measurement~\cite{Chen2024SpatialVLMEV, liao2024reasoning, shiri2024empirical}. When VLMs encode surgical images, their embeddings cannot extract quantities such as the degree of attachment or separation between structures. SurgGraph's core task, measuring \emph{how much} two structures are attached or separated, is precisely this kind of quantitative reasoning. With SurgGraph, the clinical relation (e.g., ``attachment is physical contact at consistent depth'') are translated into a geometric criterion (shared boundary pixels with low depth variance), and implements it as a deterministic function. The result is exact, reproducible, and requires no spatial reasoning training data. The full process for creating new geometric relations is detailed in Appendix~\ref{appendix:handbook}.

\emph{Clinical trust requires auditability.} For AI systems to be adopted in surgical training, clinicians must be able to verify that the system's reasoning aligns with clinical definitions~\cite{abgrall2024explainable}. SurgGraph's rules are directly auditable: a surgeon can inspect the attachment rule (shared boundary pixels with low depth variance) and confirm it matches the surgical definition of tissue attachment. This aligns with recent work showing that neurosymbolic approaches, where neural models handle perception and symbolic rules handle reasoning, achieve higher accuracy and full auditability in medical contexts~\cite{prenosil2025neurosymbolic}. VLMs, by contrast, offer no such transparency: their internal reasoning cannot be mapped to clinical criteria.

\emph{Limitations compared to VLMs.} Geometric rules require upfront engineering effort: each new relation must be explicitly defined and implemented, whereas a VLM can attempt novel queries without additional development. Geometric methods are also bounded by their perceptual inputs, reasoning only about what is visible in segmentation masks and depth maps. Relations that depend on texture (e.g., inflammation) fall outside their scope.

\subsection{Generalizability}

SurgGraph's generalizability operates at two layers. The \emph{representation
layer} concerns applying or extending the scene-graph pipeline itself, and the
\emph{application layer} concerns building applications on top of already
computed scene graphs. We provide a detailed handbook for the former in
Appendix~\ref{appendix:handbook}; the latter requires no pipeline work at all.

\emph{Representation layer.} The relation algorithms and segmentation vocabulary remain unchanged across cholecystectomy cases, but the algorithm parameters are sensitive to imaging conditions. Different laparoscopic camera systems vary in field of view, resolution, and optical characteristics, which affect pixel-level measurements such as edge proximity thresholds and depth variance. Recordings from older laparoscopic systems differ from modern ones in resolution, color profile, and noise level, which can degrade both segmentation quality and depth estimation accuracy, requiring more aggressive parameter tuning. To adapt to a new imaging setup, a developer selects 5--10 calibration frames with known ground-truth relations, runs the algorithms with default parameters, and iteratively adjusts the small number of thresholds ($\tau$, $r$, $\delta$) until outputs align with expert judgment. See Appendix~\ref{appendix:handbook}, Step~4 for the calibration procedure.

SurgGraph's four relation types --- attachment, occlusion, separation, tool action --- and their underlying geometric primitives (boundary overlap, depth comparison, convex hull gap, tooltip proximity) are broadly applicable beyond cholecystectomy. When a new procedure requires an additional relation, a developer can add it by defining the relation clinically, translating it into a geometric criterion over segmentation and depth, and implementing it as a self-contained module that integrates without modifying existing algorithms. For example, a \emph{coverage} relation for hernia repair could measure how completely a mesh covers the defect as the mask overlap ratio $|\text{mask}_{\text{mesh}} \cap \text{mask}_{\text{defect}}| / |\text{mask}_{\text{defect}}|$, verified by depth ordering. A new procedure also requires updating the segment vocabulary and the tool--action--target dictionary. See Appendix~\ref{appendix:handbook}, Steps~1--3 for the full process.

\emph{Application layer: new applications from existing scene graphs.} The
scene graphs are application-agnostic: once computed for a procedure, what
they support depends entirely on how a user queries them, and SurgGraphQA is
one such application rather than the only one. Within education, expanding the
exercise bank requires only authoring a new template, which consists of a
textual stem, scene-graph retrieval conditions, and feedback rules
(Section~\ref{sec:platform}); The exercises in this paper instantiate three templates across two learning objectives. Instructors can extend this set to new landmarks, tools, or assessment criteria using the same recordings without modifying the scene graph generation pipeline, but must author new templates grounded in the corresponding medical definitions, just like the exposure or the CVS. The same scene graphs equally support other uses: navigating a
recording by jumping to frames where a queried condition holds, summarizing a
video by sampling representative frames from each procedural stage, or
tracking relation values over time to monitor procedural progress. This
separation is what makes the representation reusable: the expertise-heavy
step of building scene graphs is done once per procedure, while applications
iterate freely on top of it.

\subsection{Limitations and Future Work}\label{sec:limitations}
First, the creation of geometric programs should move towards automation. As LLMs become increasingly capable of composing programs~\cite{gupta2023visual, suris2023vipergpt}, prompting them to decompose clinically meaningful relations into geometric properties of segmentation masks and depth maps is a promising direction.

Second, currently each frame is processed independently. Incorporating temporal models over frame-level scene graphs could smooth transient segmentation errors and enable new relation types that require multi-frame reasoning, such as tracking tooltip trajectories to infer tool force, direction of manipulation, and instrument kinematics.

Third, the current tool action target dictionary assumes fixed instrument to action mappings. Future work could learn from observed tool usage patterns to accommodate creative tool use (e.g., a grasper for blunt dissection) and capture manipulation dynamics beyond single-frame contact detection.

\section{Conclusion}

We presented SurgGraph, a training-free pipeline that generates quantitative scene graphs from laparoscopic video by translating clinical definitions into deterministic geometric programs over segmentation and depth maps. The resulting $\langle$subject, verb, object, quantity$\rangle$ representation enables retrieval by surgical query, explainable criterion-level reasoning, and automatic visual cue labeling. Technical evaluations show SurgGraph substantially outperforms state-of-the-art surgical VLMs, and a user study of SurgGraphQA, a proof-of-concept educational application of the representation, shows significant learning gains among 17 medical students, with qualitative feedback confirming the educational value of labeled images, diverse stage comparisons, and structured checklists.

\begin{acks}
This work was supported by the National Science Foundation under Grant No. 2406218, ``Multimodal Techniques to Enhance Intra- and Post-operative Learning and Coordination between Attending and Resident Surgeons.''
\end{acks}
\bibliographystyle{ACM-Reference-Format}
\bibliography{sample-base}

\appendix
\section{Lap Chole Steps}
\begin{table}[h]
\centering
\caption{Knowledge that can be learned from laparoscopic cholecystectomy video, organized by surgical step.}
\label{tab:knowledge-from-video}
\resizebox{\columnwidth}{!}{%
\begin{tabular}{p{7cm}ccc}
\toprule
Surgical Step & Anatomy & Tool & Procedure Decision Making \\
\midrule
Dissection and Exposure: The surgeon lifts the gallbladder and dissects adhesions to expose the Triangle of Calot (formed by the cystic duct, cystic artery, and common hepatic duct). & \checkmark & \checkmark & \checkmark \\
\midrule
Critical View of Safety (CVS): A standardized method is used to identify the cystic duct and artery, ensuring they are separated from surrounding structures to avoid bile duct injury. & \checkmark & \checkmark & \checkmark \\
\midrule
Clipping and Dividing: The cystic duct and cystic artery are clipped and cut using an endoscopic clip applier and scissors or cautery device. & & \checkmark & \\
\midrule
Dissecting the Gallbladder: The gallbladder is carefully removed from its attachment to the liver bed using electrocautery or an ultrasonic energy device. & & \checkmark & \\
\midrule
Removal and Closure: The gallbladder is extracted through a trocar port, the surgical site is inspected, and port sites are closed. & & \checkmark & \\
\bottomrule
\end{tabular}%
}
\end{table}
\section{Tool Usage}
\begin{table}[h]
\centering
\caption{Surgical tools, their actions, and possible targets in laparoscopic cholecystectomy.}
\label{tab:tool-action-target}
\resizebox{\columnwidth}{!}{%
\begin{tabular}{lll}
\toprule
\textbf{Tool} & \textbf{Action} & \textbf{Possible Targets} \\
\midrule
\multirow{2}{*}{Grasper}  & retract & gallbladder, liver, neck, fat, connective tissue \\
                           & grasp   & cystic\_plate, bag \\
\midrule
\multirow{2}{*}{Hook}     & dissect   & gallbladder, cystic\_artery, cystic\_duct, cystic\_plate, connective tissue, neck \\
                           & coagulate & gallbladder, cystic\_duct, cystic\_plate, liver \\
\midrule
\multirow{4}{*}{Bipolar}  & coagulate & liver, gallbladder, cystic\_artery, neck, blood, fat \\
                           & dissect   & cystic\_plate \\
                           & retract   & liver \\
                           & grasp     & bag \\
\midrule
Clipper                   & clip      & cystic\_artery, cystic\_duct, blood \\
\midrule
Scissors                  & cut       & cystic\_artery, cystic\_duct, connective tissue \\
\midrule
\multirow{3}{*}{Suction}  & aspirate  & fluid \\
                           & dissect   & gallbladder, cystic\_plate, neck \\
                           & retract   & liver \\
\bottomrule
\end{tabular}%
}
\end{table}

\section{Phase Recognition Rule Set}\label{appendix:phase-rules}

Each surgical phase is classified using a conjunction of scene graph conditions:

\begin{itemize}[leftmargin=*]
\item \textbf{Preparation:} $\langle$fat/intestine/connective tissue, attach/occlude, gall/neck$\rangle > 20\%$
\item \textbf{Calot Triangle Dissection:} $\langle$fat/intestine/connective tissue, attach/occlude, gall/neck$\rangle < 20\%$ AND ($\langle$cystic duct, separate, cystic artery$\rangle < 80\%$ OR $\langle$gallbladder, attach, cystic plate$\rangle > 80\%$)
\item \textbf{Clipping/Cutting:} $\langle$cystic duct, separate, cystic artery$\rangle > 80\%$ AND $\langle$gallbladder, attach, cystic plate$\rangle < 80\%$ AND presence of clip applier or scissors
\item \textbf{Gallbladder Dissection:} no cystic duct/artery present AND ($\langle$gall, attach, liver$\rangle > 0$ OR $\langle$gall, attach, cystic plate$\rangle \in [20\%, 80\%]$)
\item \textbf{Gallbladder Retraction:} $\langle$grasper, retract, gall$\rangle$ AND $\langle$gall, attach, liver$\rangle = 0$ AND $\langle$gall, attach, cystic plate$\rangle = 0$
\item \textbf{Cleaning/Coagulation:} $\langle$suction, suct, blood$\rangle$ OR $\langle$hook, coagulate, liver$\rangle$
\item \textbf{Gallbladder Packaging:} $\langle$bag, area, bag$\rangle > 0$
\end{itemize}

\section{Exercise Templates}\label{appendix:templates}
\begin{figure}
  \centering
  \includegraphics[width=0.5\textwidth]{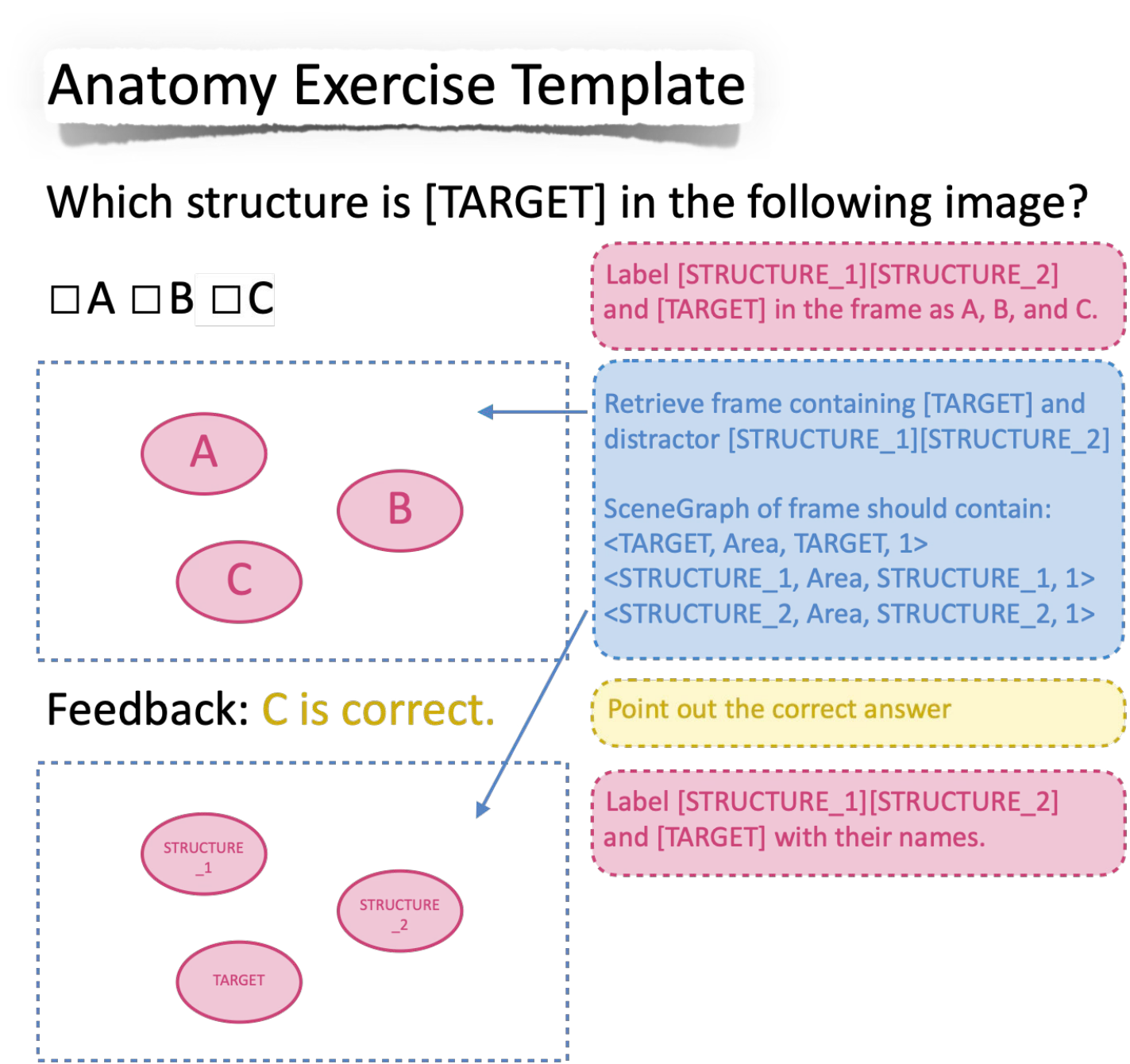}
  \caption{Exercise template for anatomy recognition. Blue: image retrieval via scene graph query. Yellow: generated textual feedback. Pink: visual cue labeling on images.}
  \Description{}
  \label{fig:template-anatomy}
\end{figure}

\begin{figure}
  \centering
  \includegraphics[width=0.5\textwidth]{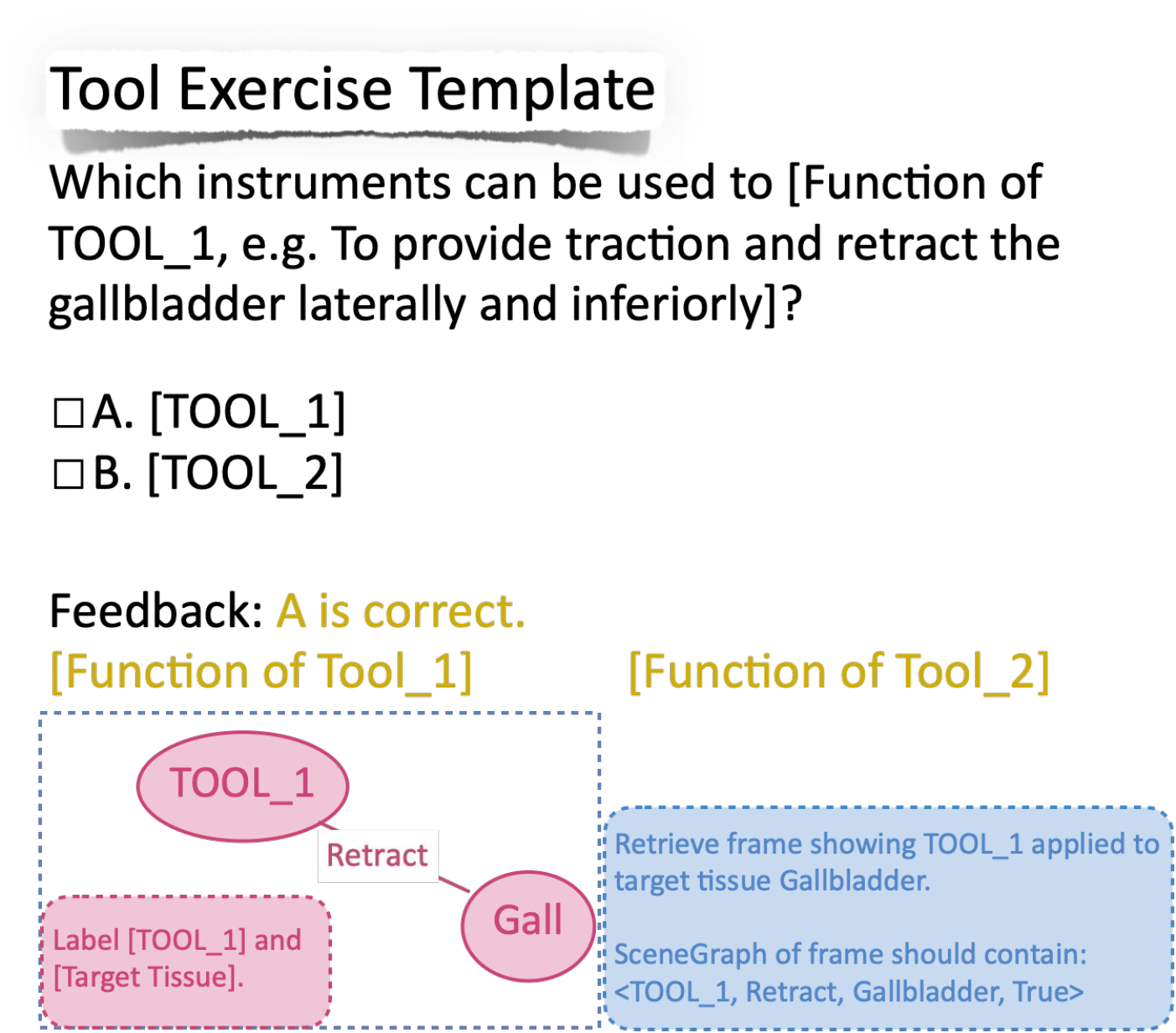}
  \caption{Exercise template for tool usage. Blue: image retrieval via scene graph query. Yellow: generated textual feedback. Pink: visual cue labeling on images.}
  \Description{}
  \label{fig:template-tool}
\end{figure}

Figures~\ref{fig:template-anatomy} and~\ref{fig:template-tool} show the
anatomy-recognition and tool-usage templates; the procedural-assessment
template appears in the main text (Figure~\ref{fig:template-mcq}).

\section{Mode B: Diverse Stage Retrieval}\label{appendix:modeb}

Mode B (Section~\ref{sec:retrieval}) returns a small pool of frames that span different stages of completion of a learning objective, rather than the single best match returned by Mode A. Each learning objective is expressed as $d$ relation-level requirements (for exposure, $d=14$; Figure~\ref{fig:retrieval}), which the feedback layer composes into the textbook criteria of Section~\ref{sec:explanation}. Let $\mathbf{r}\in\{0,1\}^d$ be the target vector of the objective and $\mathbf{f}_t\in[0,1]^d$ the normalized scene-graph values of frame $t$. Mode B proceeds in four steps:

\begin{enumerate}[leftmargin=*]
\item \textbf{Requirement evaluation.} For every frame $t$ and requirement $i$, compute a satisfaction bit $s_{t,i} = \mathbb{1}[\,|f_{t,i} - r_i| \le 0.2\,]$. The tolerance of 0.2 is the ``largely absent/present'' landmark of Section~\ref{sec:regularization}; boolean requirements such as tool actions are compared exactly. Frames in which none of the structures named by the objective is present (all relevant area values are 0) are discarded, since they carry no information about the objective.
\item \textbf{Pattern grouping.} The binary vector $\mathbf{s}_t\in\{0,1\}^d$ is the frame's \emph{satisfaction pattern}. Frames with identical patterns are placed in the same group, so each group corresponds to one distinct combination of met and unmet requirements (e.g., ``only fat attachment unmet'' vs.\ ``fat attachment and grasper retraction both unmet''). Two visually different frames from different moments of the procedure fall into the same group if the same requirements remain unmet.
\item \textbf{Group ordering.} Groups are ordered by their number of unmet requirements, $\|\mathbf{r}-\mathbf{s}_t\|_1$, from most unmet to fully met. Because relation values evolve with procedural progress, this ordering follows the procedure: in the video of Figure~\ref{fig:retrieval}, five groups emerge with 3, 2, 2, 1, and 0 of 14 requirements unmet.
\item \textbf{Sampling.} One frame is drawn uniformly at random from each group (or $k$ frames per group when a template needs more options). The result is a pool of frames, one per stage of completion, from which exercise templates draw their options: two-option procedural exercises take the correct option from the fully met group and the distractor from a group with at least one unmet requirement, while open-ended exercises take a single frame from any group. Each sampled frame's pattern $\mathbf{s}_t$ directly yields its checklist feedback (Section~\ref{sec:explanation}).
\end{enumerate}

\section{Handbook: Defining Relations for a New Surgery}\label{appendix:handbook}

This appendix provides a step-by-step guide for extending SurgGraph to a new surgical procedure. The process requires collaboration between a surgical domain expert and a system developer, and involves four stages.

\subsection{Step 1: Prepare Perceptual Inputs}

The foundation of any geometric relation is accurate segmentation and depth estimation. To set up the perceptual pipeline for a new procedure:

\begin{enumerate}[leftmargin=*]
\item \textbf{Processing input maps.} Select frames from the new procedure that cover all anatomical structures, tools, and procedural stages of interest. For a typical 2-hour surgical video, approximately 50 annotated frames are sufficient. A surgical domain expert labels the segmentation classes on these frames, then SAM3 (or a comparable segmentation model) propagates the labels across the full video via tracking. Depth maps can be obtained directly from a monocular depth estimation model (e.g., Depth Anything) with no manual annotation required; values should be normalized to $[0, 1]$.
\end{enumerate}

\subsection{Step 2: Define Relations as Geometric Rules}

For each clinically meaningful relation in the new procedure, follow the three-step translation process:

\begin{enumerate}[leftmargin=*]
\item \textbf{Clinical definition.} The surgical expert defines the relation in clinical terms. For example: ``the mesh \emph{covers} the defect when it extends beyond the defect margins,'' or ``the ureter is \emph{exposed} when it is visible and separated from surrounding tissue.''
\item \textbf{Geometric translation.} Translate the clinical definition into a computable criterion over segmentation masks and depth maps. Common geometric primitives include:
\begin{itemize}
\item \emph{Boundary overlap:} shared boundary length between two masks (used for attachment).
\item \emph{Depth comparison:} relative depth at shared boundaries (used to distinguish attachment from occlusion).
\item \emph{Area ratio:} overlap area between two masks normalized by one mask's total area (used for coverage).
\item \emph{Convex hull gap:} distance between convex hulls of two masks (used for separation).
\item \emph{Proximity at functional region:} depth and mask overlap within a tool's tip region (used for tool--tissue interaction).
\end{itemize}
\item \textbf{Output format and normalization.} Each relation should specify its output type: continuous numeric values (e.g., shared boundary length), boolean flags (e.g., tool--tissue contact detected), or vector values (e.g., spatial direction of separation). Because laparoscopic camera movement causes apparent size changes in all structures, raw pixel-based measurements must be normalized. Identify a \emph{reference structure} whose physical size remains approximately constant throughout the procedure (e.g., the gallbladder in cholecystectomy), and divide relation values by that structure's pixel area in each frame. This per-entity normalization compensates for camera zoom and distance. Additionally, divide each pairwise relation by its maximum value across all frames in a video to scale values to $[0, 1]$, making them comparable across relation types and across videos with different imaging conditions.
\item \textbf{Implementation.} Implement the criterion as a deterministic function that takes a frame's segmentation mask and depth map as input and returns a normalized value for each relevant entity pair. The function should be self-contained and stateless, operating independently of other relation modules.
\end{enumerate}

\subsection{Step 3: Build the Tool--Action--Target Dictionary}

Tool usage relations require a dictionary that maps each surgical instrument to its possible actions and anatomical targets. To construct this for a new procedure:

\begin{enumerate}[leftmargin=*]
\item \textbf{Consult surgical textbooks and operative manuals} to enumerate the instruments used in the procedure, the actions each instrument performs (e.g., grasp, cut, coagulate, retract, clip), and the anatomical structures each action targets.
\item \textbf{Organize into a structured table} of $\langle$tool, action, target$\rangle$ triplets (see Table~\ref{tab:tool-action-target} for the laparoscopic cholecystectomy example).
\item \textbf{Validate with the surgical expert} to ensure completeness and correctness. The dictionary constrains which tool--tissue interactions the system recognizes; missing entries will result in undetected relations.
\end{enumerate}

This dictionary serves as the knowledge base for tool action detection: when the geometric algorithm detects a tool tip in contact with a tissue mask, it looks up the valid $\langle$tool, action, target$\rangle$ triplets to determine the relation type.

\subsection{Step 4: Calibrate Parameters on Sample Frames}

Each geometric rule has a small number of tunable parameters (e.g., depth variance thresholds, proximity radii). To calibrate these for a new procedure:

\begin{enumerate}[leftmargin=*]
\item \textbf{Select calibration samples.} Choose 5--10 frames where the ground-truth relation value is known or can be estimated by the surgical expert (e.g., frames where two structures are clearly attached vs.\ clearly separated).
\item \textbf{Run the relation algorithm} with default parameters and compare outputs to the expert's assessments. Identify cases where the algorithm over- or under-estimates the relation value.
\item \textbf{Adjust parameters iteratively.} Tune thresholds (e.g., depth variance $\tau$, proximity radius $r$) to align algorithmic outputs with expert judgment on the calibration samples. Because each relation has only 2--3 parameters, this can typically be done by manual inspection rather than automated optimization.
\item \textbf{Validate on held-out frames.} Test the calibrated parameters on a separate set of frames to verify generalization. If the procedure uses a different laparoscopic camera or imaging system, parameters may need re-calibration, as field of view, resolution, and depth scale can differ across devices.
\end{enumerate}

Following these four steps, a new procedure can be integrated into the SurgGraph pipeline, producing structured scene graphs that enable the same retrieval, labeling, and educational feedback capabilities demonstrated for laparoscopic cholecystectomy.

\section{Participant Demographics}\label{appendix:demographics}
\begin{table}[h]
\centering
\caption{Participant demographics (N=19). Familiarity with laparoscopic cholecystectomy is self-reported on a 1--5 scale.}
\label{tab:demographics}
\resizebox{\columnwidth}{!}{%
\begin{tabular}{llllll}
\toprule
\textbf{ID} & \textbf{Gender} & \textbf{Race/Ethnicity} & \textbf{Training Level} & \textbf{Surgical Interest} & \textbf{Familiarity} \\
\midrule
P1  & Female & Asian                        & M4       & Yes & 3 \\
P2  & Female & Middle Eastern/North African  & M3       & No  & 3 \\
P3  & Female & Asian                        & M4       & Yes & 3 \\
P4  & Female & Black/African American       & M2       & Yes & 2 \\
P5  & Male   & Black/African American       & M1       & Yes & 2 \\
P6  & Female & Multiracial                  & M4       & Yes & 3 \\
P7  & Female & Middle Eastern/North African  & M3       & Yes & 2 \\
P8  & Male   & Black/African American       & M1       & Yes & 1 \\
P9  & Female & Asian                        & M3       & No  & 3 \\
P10 & Female & Asian                        & M3       & Yes & 3 \\
P11 & Female & Black/African American       & Resident & Yes & 4 \\
P12 & Male   & White                        & M3       & Yes & 3 \\
P13 & Male   & Asian, White, Multiracial    & M3       & No  & 3 \\
P14 & Male   & White                        & M3       & Yes & 3 \\
P15 & Female & Asian                        & M4       & No  & 3 \\
P16 & Female & White                        & Resident & Yes & 5 \\
P17 & Male   & Asian, White, Multiracial    & M2       & Yes & 3 \\
P18 & Female & Asian                        & M3       & No  & 2 \\
P19 & Male   & Asian                        & M2       & Yes & 2 \\
\bottomrule
\end{tabular}%
}
\end{table}

\section{Separation IoU Validation}\label{appendix:separation-iou}
\begin{figure}[h]
  \centering
  \includegraphics[width=\columnwidth]{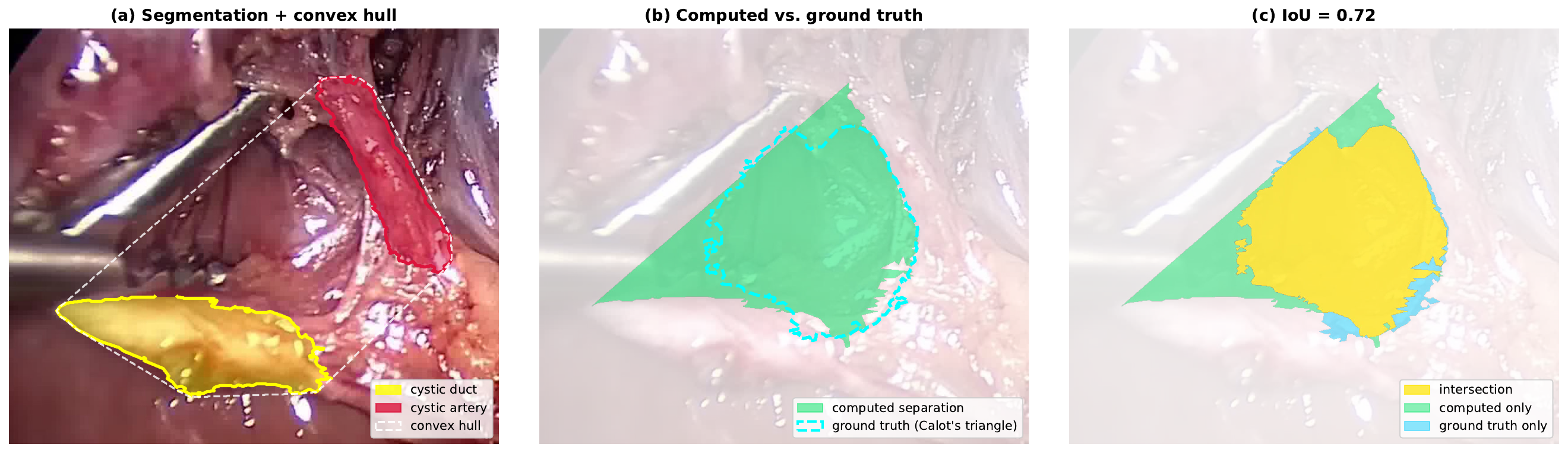}
  \caption{Separation IoU validation. (a) Segmentation of cystic duct and cystic artery with their joint convex hull. (b) SurgGraph's computed in-between area (green), after excluding gallbladder and neck regions, overlaid with expert-labeled Calot's triangle ground truth (cyan outline). (c) IoU decomposition showing intersection (yellow), computed-only (green), and ground-truth-only (cyan) regions.}
  \Description{}
  \label{fig:separation-iou}
\end{figure}

\section{Segment Categories}\label{appendix:segments}
SurgGraph segments 19 categories: \emph{Anatomical structures:} fat, liver, intestine, gallbladder, neck, cystic duct, cystic artery, cystic plate, blood. \emph{Tools:} grasper, L-hook, scissors, dissector, clip applicator, suction. \emph{Other:} background, wall, unknown, bag.

\section{CVS Feedback Criteria}\label{appendix:cvs-criteria}
The CVS criteria and their scene graph mappings for explanation generation:
\begin{itemize}[leftmargin=*]
\item $\langle$cystic duct, separate, cystic artery, $>$0.6$\rangle$ $\rightarrow$ ``Cystic duct and artery clearly separated and skeletonized; the hepatocystic triangle is clear of all fibrofatty tissue''
\item $\langle$cystic artery, area, $>$0$\rangle$ $\rightarrow$ ``Cystic artery is visible''
\item $\langle$cystic duct, area, $>$0$\rangle$ $\rightarrow$ ``Cystic duct is visible''
\item $\langle$gallbladder, attach, cystic plate, $<$0.2$\rangle$ $\rightarrow$ ``Lower third of the gallbladder dissected off the liver bed''
\end{itemize}

\section{Pretest/Posttest Questions and Rubrics}\label{appendix:rubrics}
The pretest and posttest questions used to measure learning gain, with answer rubrics. Section 1 covers Exposure; Section 2 covers CVS.

\begin{figure*}[h]
\centering
\includegraphics[width=0.24\textwidth]{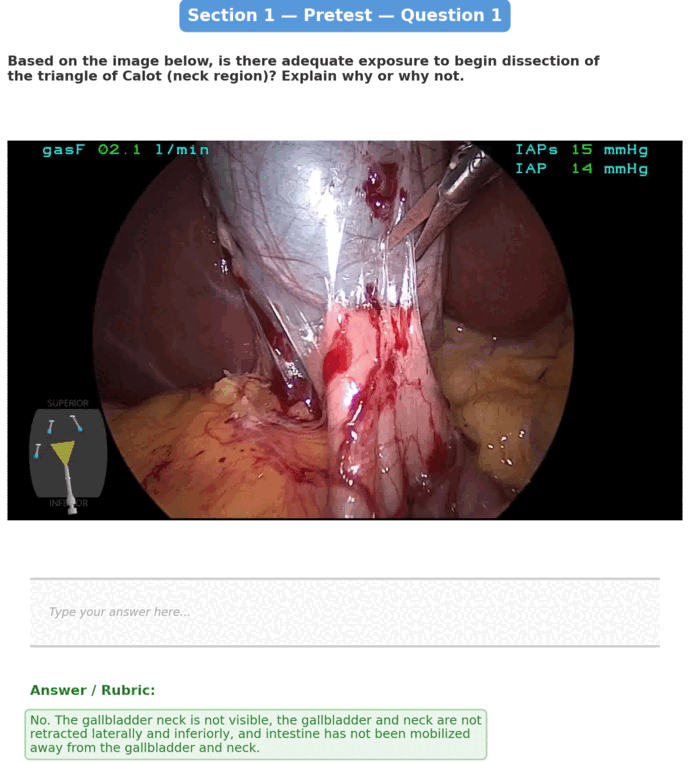}
\includegraphics[width=0.24\textwidth]{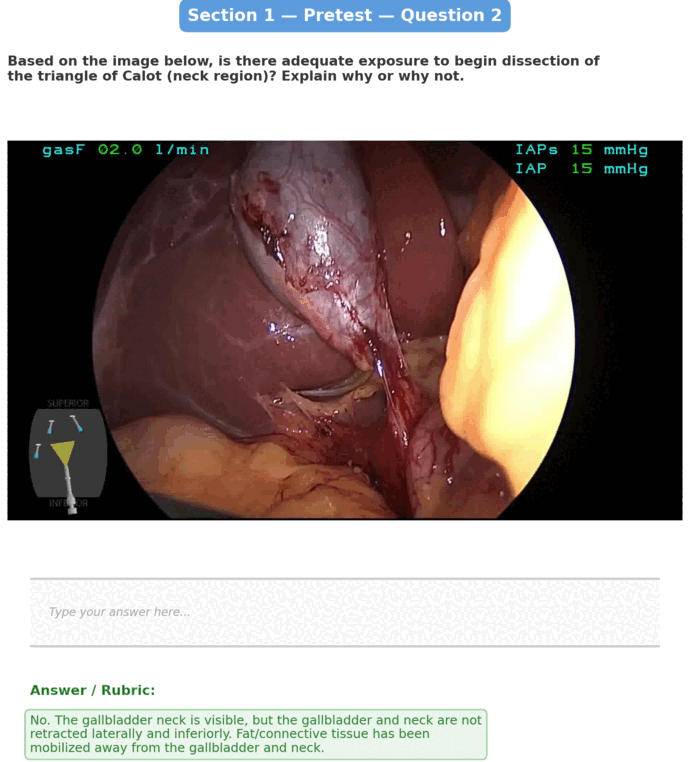}
\includegraphics[width=0.24\textwidth]{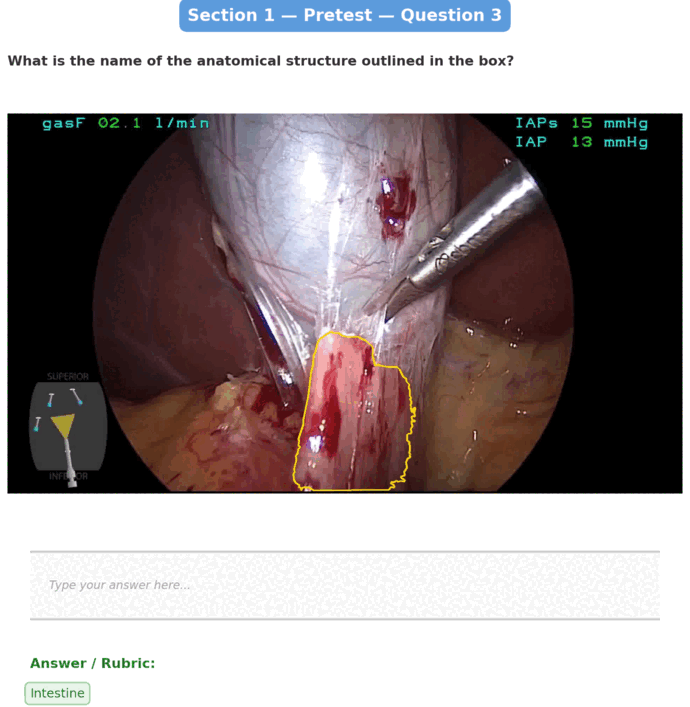}
\includegraphics[width=0.24\textwidth]{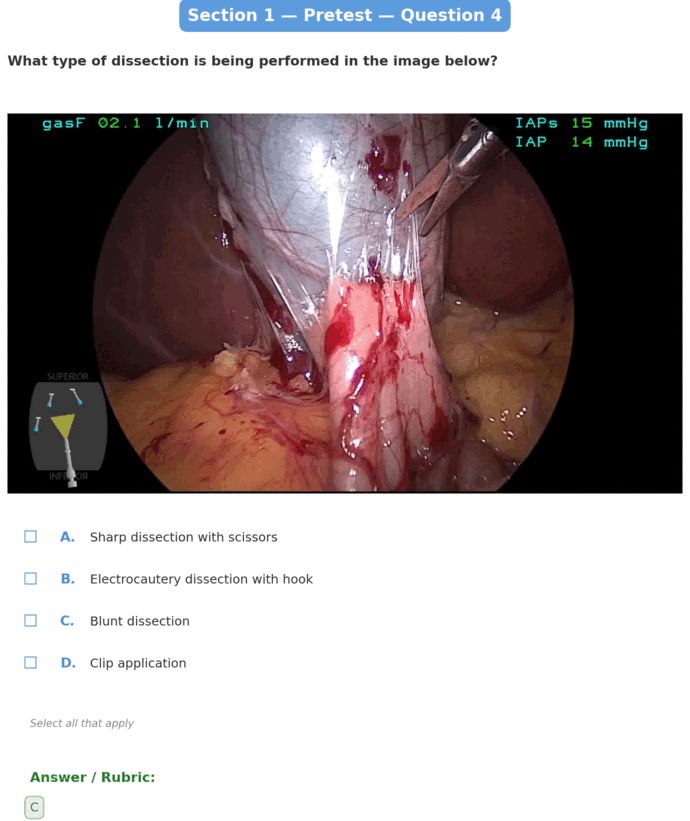}
\caption{Section 1 (Exposure): Pretest questions 1--4 with rubrics.}
\Description{}
\label{fig:pretest-s1}
\end{figure*}

\begin{figure*}[h]
\centering
\includegraphics[width=0.24\textwidth]{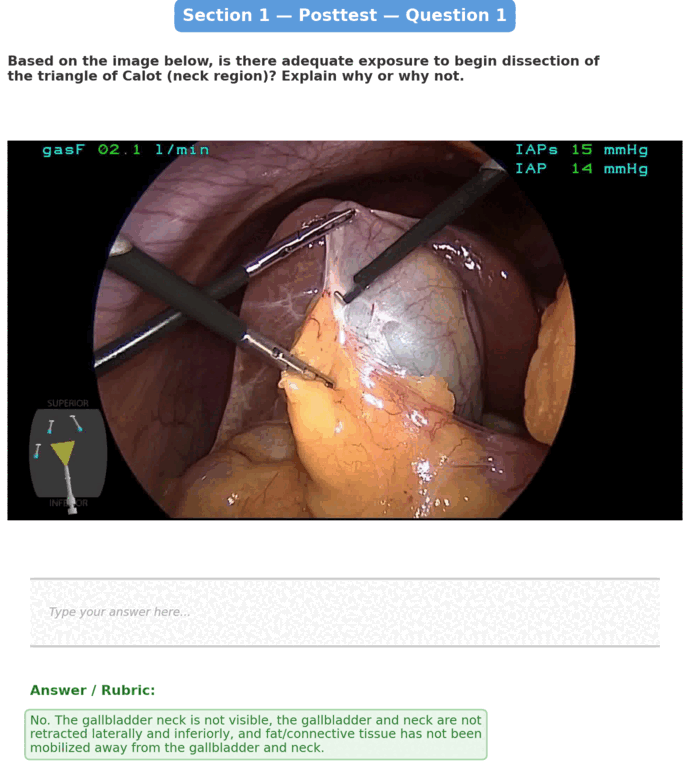}
\includegraphics[width=0.24\textwidth]{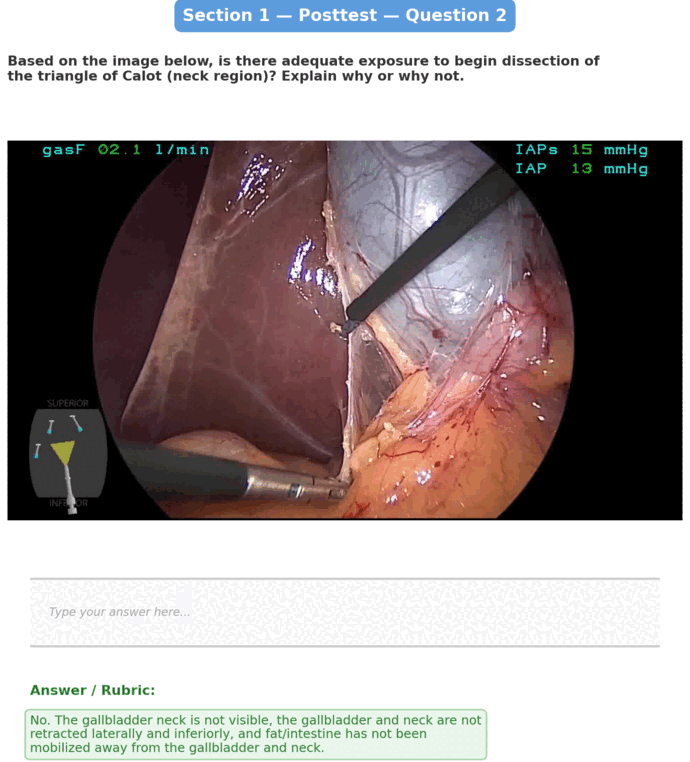}
\includegraphics[width=0.24\textwidth]{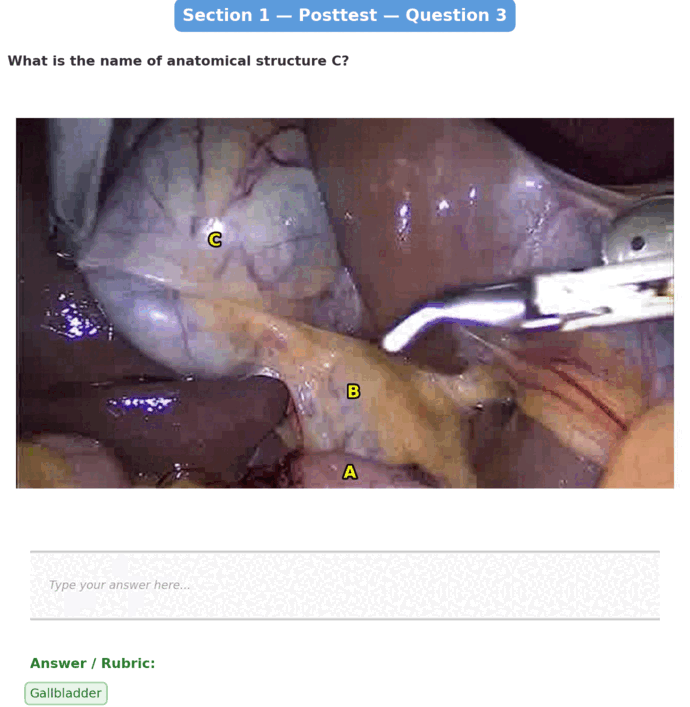}
\includegraphics[width=0.24\textwidth]{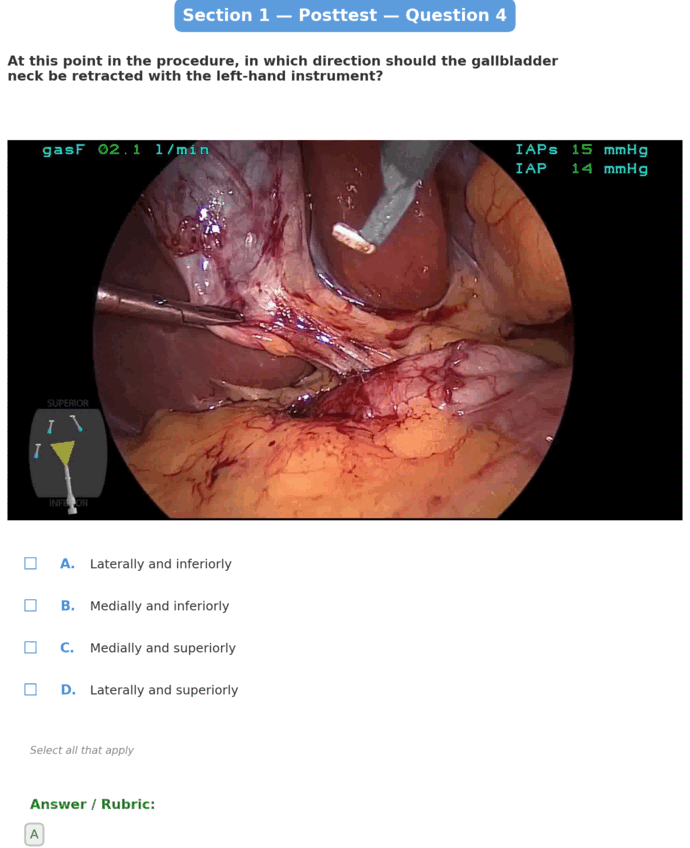}
\caption{Section 1 (Exposure): Posttest questions 1--4 with rubrics.}
\Description{}
\label{fig:posttest-s1}
\end{figure*}

\begin{figure*}[h]
\centering
\includegraphics[width=0.24\textwidth]{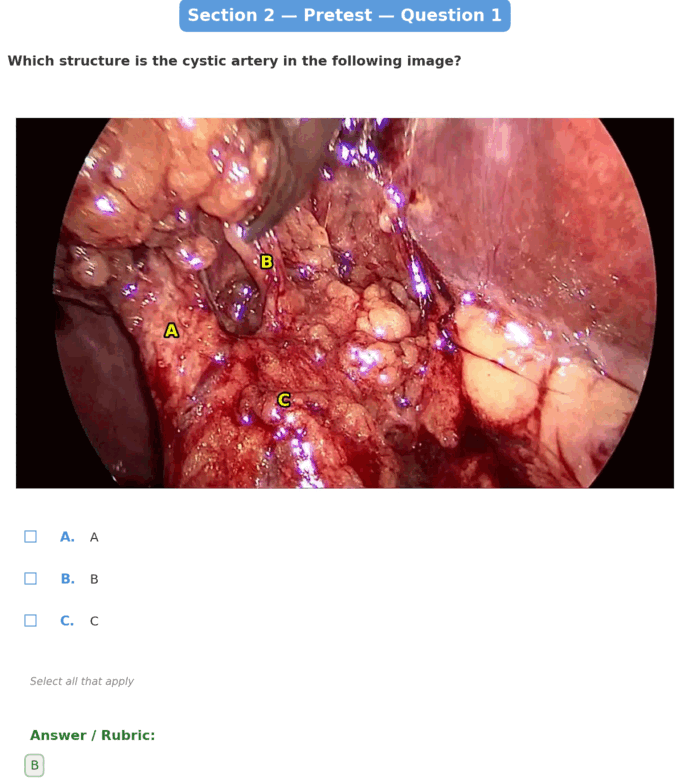}
\includegraphics[width=0.24\textwidth]{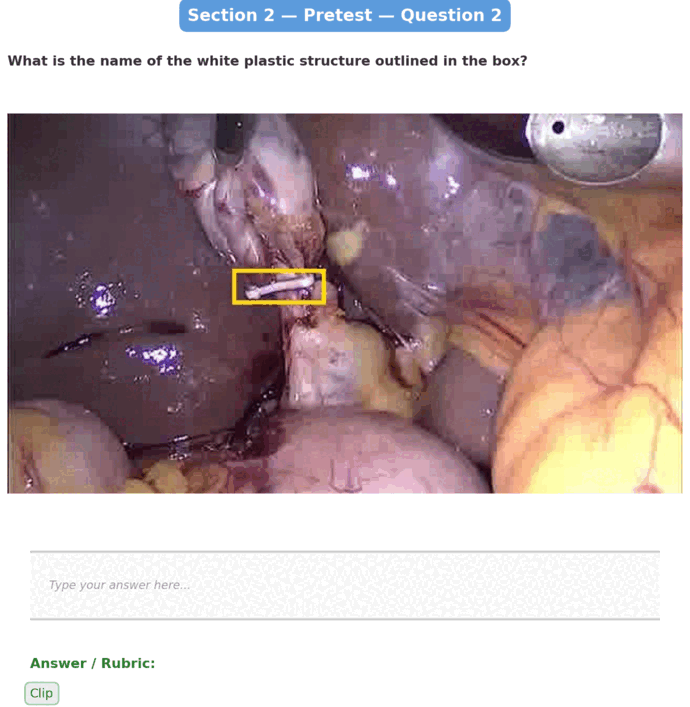}
\includegraphics[width=0.24\textwidth]{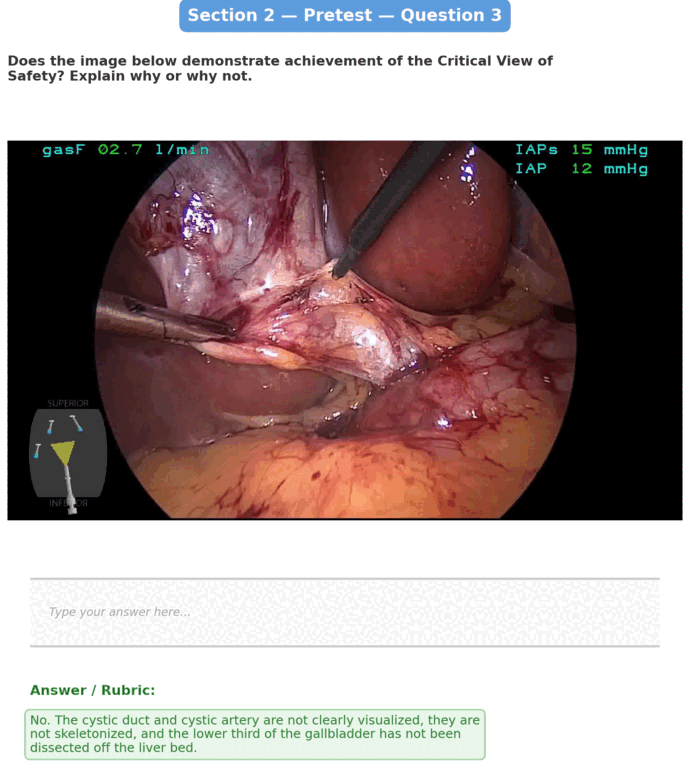}
\includegraphics[width=0.24\textwidth]{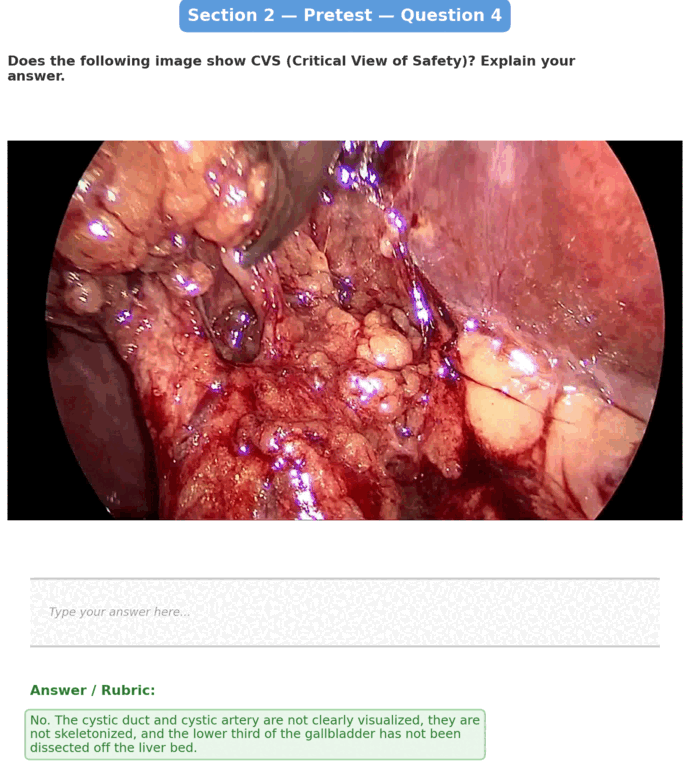}
\caption{Section 2 (CVS): Pretest questions 1--4 with rubrics.}
\Description{}
\label{fig:pretest-s2}
\end{figure*}

\begin{figure*}[h]
\centering
\includegraphics[width=0.24\textwidth]{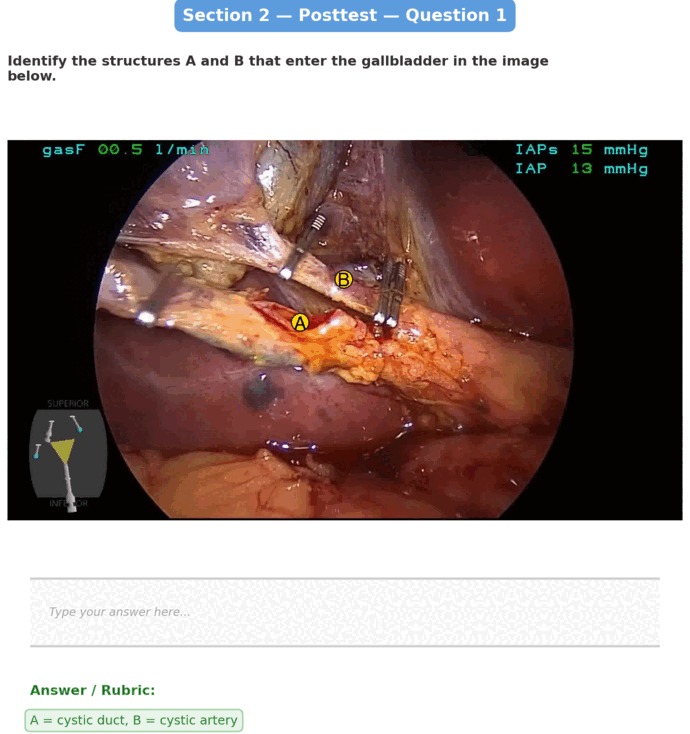}
\includegraphics[width=0.24\textwidth]{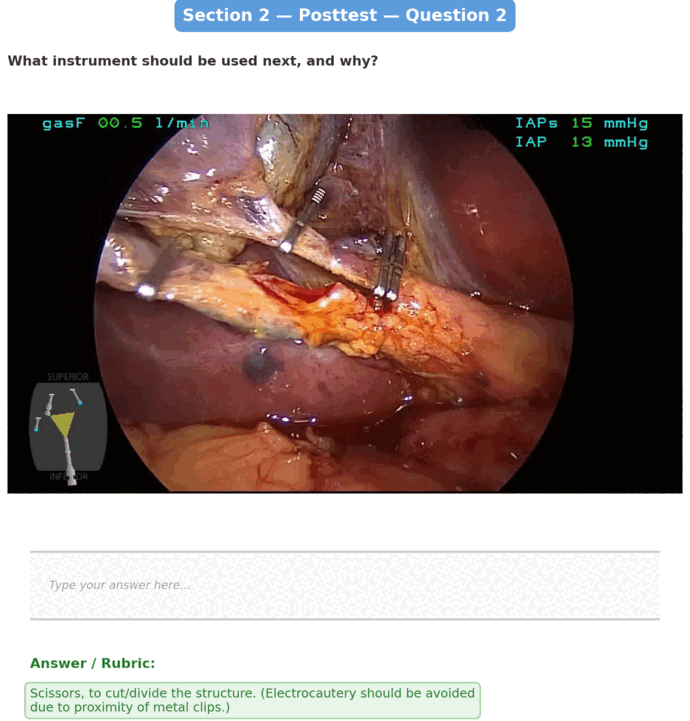}
\includegraphics[width=0.24\textwidth]{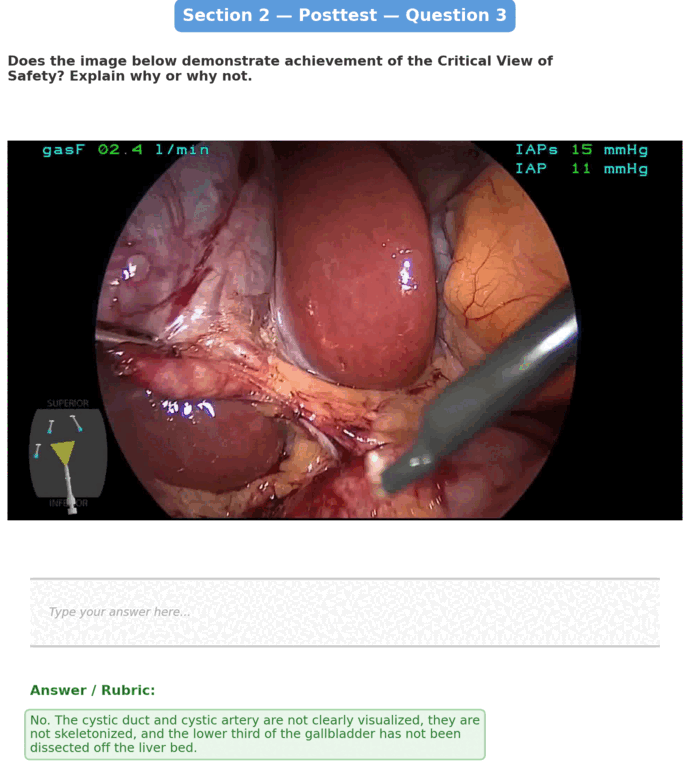}
\includegraphics[width=0.24\textwidth]{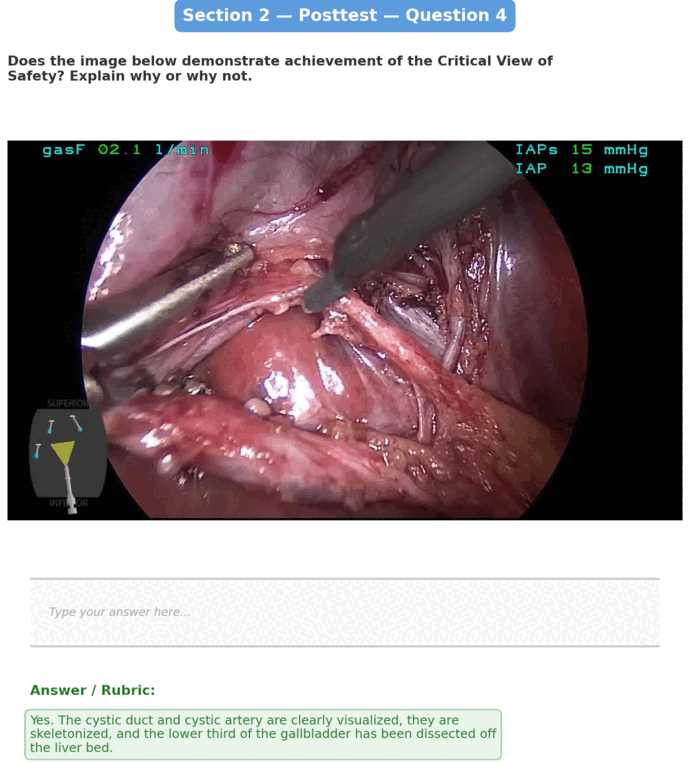}
\caption{Section 2 (CVS): Posttest questions 1--4 with rubrics.}
\Description{}
\label{fig:posttest-s2}
\end{figure*}

\end{document}